%% file: paper_main.tex
\documentclass[onecolumn]{article}
\usepackage{geometry}
\usepackage{amsmath,amssymb,amstext}
\usepackage[english]{babel}
\usepackage{enumerate}
\usepackage[T1]{fontenc}
\usepackage{url}
\usepackage{multirow}
\usepackage{graphicx}
\usepackage{cuted}
\usepackage{isotope}
\usepackage{newtxtext,newtxmath}
\usepackage{epstopdf}
\usepackage{tabularx}
\usepackage{booktabs}
\usepackage{color}
\usepackage{siunitx}
\usepackage[hidelinks]{hyperref}
\usepackage[biblabel]{cite}
\usepackage[capitalize]{cleveref}
\usepackage{braket}
\usepackage{gensymb}
\usepackage{subcaption}
\usepackage[utf8]{inputenc}
\usepackage[version = 4]{mhchem}
\usepackage[compact]{titlesec}

\graphicspath{{../Figures/}}

\titleformat{\section}[block]
{\normalfont\bfseries}
{\thesection}{\compact}{}

\begin{document}

    \title{Tunable magnons in a dual-gated 2D antiferromagnet}
    \author{Nele Stetzuhn$^{1,2}$, Abhijeet M. Kumar$^{1}$, Sviatoslav Kovalchuk$^{1}$, Denis Yagodkin$^{1}$,\\ Louis Simon$^{1}$, Samuel Mañas-Valero$^{3,4}$, Eugenio Coronado$^{3}$, Takashi Taniguchi$^{6}$, \\Kenji Watanabe$^{6}$, Deepika Gill$^{2}$, Sangeeta Sharma$^{1,2}$, Piet Brouwer$^{1}$,\\ Clemens von Korff Schmising$^{2,5}$, Stefan Eisebitt$^{2,5}$, Kirill I. Bolotin$^{1}$}
    \date{%
    \small
    $^1$ Department of Physics, Freie Universität Berlin, Arnimallee 14, 14195 Berlin, Germany\\%
    $^2$ Max Born Institute for Nonlinar Optics and Short-Pulse Spectroscopy, Max-Born-Str. 2A, 12489 Berlin, Germany\\
    $^3$ Instituto de Ciencia Molecular, Universidad de Valencia, Dr. Moliner 50, Burjasot, 46100, Spain\\
    $^4$ Department of Quantum Nanoscience, Kavli Institute of Nanoscience, Delft University of Technology, Delft 2628CJ, the Netherlands \\
    $^5$ Institut für Optik und Atomare Physik, Technische Universität Berlin, Straße des 17. Juni 135, 10623 Berlin, Germany \\
    $^6$ National Institute for Materials Science, Namiki 1-1, Tsukuba, 305-0044, Ibaraki, Japan \\[2ex]%
    \normalsize
}

\maketitle
\input{abstract}

\input{Introduction}
\input{results}
\input{discussion}
\input{main}
\newpage
\input{methods}
\bibliography{sources}
\bibliographystyle{unsrt}
\input{figures}

\include{main_si}

\end{document}

%% file: abstract.tex
\textit{
The layered antiferromagnet CrSBr features magnons coupled to other quasiparticles, including excitons and polaritons, enabling their easy optical accessibility. In this work, we investigate the tunability of magnons in few-layered devices in response to changes in carrier density and the application of a perpendicular electric field. We demonstrate an on-chip tunability of the in- and out-of-phase magnon frequencies by up to 2 GHz. While the frequencies of both modes increase with the electron density, we observe an asymmetric response with respect to the electric field in a dual-gated trilayer device. To understand the mechanism of this disparity, we develop a layer-resolved macrospin model describing the magnetic dynamics in thin, non-uniformly doped devices. Through this model we establish the doping- and electric-field-dependence of the exchange interaction, magnetic anisotropy, and magnetic moment of individual layers.
Our results advance the applications of gate-tunable magnonic devices based on 2D materials.
}

%% file: Introduction.tex
\section*{Introduction}
Two-dimensional (anti)ferromagnets have garnered attention as an easily tunable platform to study static and ultrafast magnetism. Thin van der Waals magnets, like other two-dimensional materials, react strongly to external fields \cite{Kim2021}, strain \cite{Diederich2022,Esteras2022,Pizzochero2020,Ren2021,Hu2020,Cenker2022}, intercalation \cite{Deng2018}, and the formation of Moiré potentials \cite{Xu2021,Wang2020_2,Tong2018}, enabling reversible control of their magnetic properties. 
The semiconducting layered antiferromagnet CrSBr is of special interest due to its environmental stability \cite{Telford2020,Torres2023,Ziebel2024} and strong anisotropy stemming from its orthorhombic lattice structure \cite{Klein2023}. This anisotropy results in monolayers of CrSBr being easy-axis ferromagnets along the crystallographic $\hat{b}$-axis, whereas neighboring layers couple antiferromagnetically below the Néel temperature of $\SI{132}{\kelvin}$ \cite{Lopez2022}. It also gives rise to a 1D character of the electronic bandstructure of CrSBr \cite{Klein2023}. This, in turn, influences its transport properties \cite{Wu2022,Telford2022} as well as the directionality of its tightly bound excitons \cite{Smolenski2025,Liebich2025}. \\  
Early studies on magnons in CrSBr have demonstrated their their long coherence length \cite{Bae2022}, broadband tunability \cite{Diederich2022} and strong coupling to excitons \cite{Wilson2021,Bae2022,Dirnberger2023,Wang2023,Brennan2024}. In contrast to conventional antiferromagnets, in which magnetic order is often elusive to optical probes, this coupling allows one to observe magnon oscillations in absorption-based techniques at the excitonic \cite{Bae2022,Sun2024} and polaritonic resonances \cite{Dirnberger2023}.
The majority of the studies on magnons in CrSBr have so far focused on bulk-like samples \cite{Bae2022,Dirnberger2022,Bae2024}. However, a new range of phenomena appears in thin 2D magnets due to their sensitivity to external perturbations. For instance, the carrier density and electric field in few-layer samples, controlled by external gates, influence material properties such as the coercive field \cite{Jiang2018_2}, critical temperature \cite{Wang2016, Wang2020} and magnetic ground state \cite{Ren2021,Huang2018}. While the effect of gating on magnon energies and transport has been studied in a few selected 2D magnetic systems \cite{Zhang2020_3,Qi2023,Hendriks2024,Wal2024}, the response of magnons in thin CrSBr to external electric fields remains unexplored. The detailed understanding of such a response is especially important in the context of developing spintronic applications of nanostructured 2D materials \cite{Chumak2015}. \\
Indeed, characterizing the response of a thin antiferromagnet to electrostatic gating is challenging within the framework of current models:
To describe magnetic excitations in bulk-like layered antiferromagnets, one commonly uses a macrospin model in which all spins within one layer are locked by strong intralayer exchange \cite{Gurevich1996,Rezende2019}. The complete bulk can then be described by two macrospins alternating between layers. 
In thin layered systems, however, an out-of-plane electric field breaks the symmetry between the layers, leading to inhomogeneous doping across the device. 
This leads to several questions about the detailed microscopic mechanisms governing magnon dynamics: 
How does the presence of free carriers affect the magnetic moment, interlayer exchange and magnetic anisotropies in CrSBr?
What is the role of differing magnetic properties in individual layers as a result of electric-field-induced symmetry-breaking?
Can a layer-resolved macrospin model describe the resulting magnonic behavior? Here, we address these questions by exploring the response of magnons in few-layer CrSBr devices to gate-induced carrier densities and fields. \\

%% file: results.tex
\section*{Results}
We utilize time-resolved (tr) reflectivity to investigate the gate-dependent behavior of magnons in CrSBr (see methods).
The experimental platform to control electric field and doping consists of a thin CrSBr flake ($3-8$ layers, AFM in SI section 10) sandwiched between two graphite gates, each separated from the CrSBr by few-layered hBN ($5 - \SI{11}{\nano\meter}$), as shown in \cref{fig:1}a.
The CrSBr is grounded throughout the experiment, while the gate voltages $V_{\text{b}}$ and $V_{\text{t}}$ are separately applied to bottom and top graphite, 
respectively. 
In this scheme, the sum of gate voltages $V_{\text{b}} + V_{\text{t}}$  shifts the Fermi level in CrSBr, consequently changing its doping, while the difference 
between gate voltages $\Delta V = V_{\text{t}} - V_{\text{b}}$  controls the strength of a perpendicular electric field through the device. All measurements were performed at $T=\SI{10}{\kelvin}$, well below the Néel temperature (temperature-dependent data in SI section 2).
The magnons are excited with an ultrafast laser pulse (pulse width $\sim \SI{140}{\femto\second}$) tuned to the exciton or trion resonance at $\SI{1.376}{\eV}$ or $\SI{1.350}{\eV}$, respectively, depending on the doping level of the system (SI section 1). Subsequently, we probe the tr-reflectivity response at the same wavelength. To enhance the signal strength, we apply a constant external magnetic field $H_0 \approx \SI{0.1}{\tesla}$ using a permanent magnet in proximity of the sample (SI section 1).\\
First, we explore the effect of dual-gating in a trilayer of CrSBr. 
In \cref{fig:1}b, we observe pronounced oscillations in the tr-reflectivity traces for different values of $V_{\text{b}} = V_{\text{t}}$ (orange: $\SI{-0.55}{\volt}$, purple: $\SI{5.7}{\volt}$; exponential background subtracted from both). The oscillation amplitude decreases going from low (purple) to high (orange) gate voltages, as the spectral weight between exciton and trion resonance depends on the sample doping.
The two oscillation frequencies around $\SI{19.1(1)}{\giga\hertz}$ and $\SI{28.1(1)}{\giga\hertz}$ ($V_{\text{b}} = V_{\text{t}} = \SI{-0.55}{\volt}$ in inset of \cref{fig:1}b) correspond to the in phase (macrospins in neighboring layers oscillate in phase with each other) and out of phase (macrospins in neighboring layers oscillate out of phase phase with each other) zero momentum magnon modes $f_{\text{IP}}$ and $f_{\text{OP}}$, in accordance with previous reports \cite{Cham2022,Bae2022,Bae2024,Sun2024}. As seen in \cref{fig:1}c, increasing the gate voltages leads to an upshift of the frequencies of both modes, to a maximum of $f_{\text{IP}} = \SI{21.1(1)}{\giga\hertz}$ and $f_{\text{OP}} = \SI{31.0(1)}{\giga\hertz}$ ($V_{\text{b}}$  = $V_{\text{t}} = \SI{5.7}{\volt}$, see inset of \cref{fig:1}b).
Notably, this shift is nonlinear -- in particular, below $V_{\text{t}} = V_{\text{b}} \approx \SI{1}{\volt}$, there is no significant change from the initial frequencies.
In the complete $V_{\text{b}}$-$V_{\text{t}}$-dependence of $f_{\text{IP}}$ and $f_{\text{OP}}$ shown in \cref{fig:1}d, e, we observe an asymmetry between the two modes in response to the perpendicular electric field $F_{\mathrm{z}}$ (arrows mark the direction of increase in doping $n$ or field $F_{\mathrm{z}}$). Specifically, $f_{\text{IP}}$ increases with $V_{\text{t}}$, while $f_{\text{OP}}$ reacts to changes in $V_{\text{b}}$, suggesting a surprising influence of the field direction on magnon frequencies. In the following, we develop a microscopic picture to describe these dependencies. \\
To understand the mechanism behind the gate-dependent magnon frequencies, it is crucial to examine the carrier density configuration of our device. 
Generally, the ratio of intensities between neutral and charged excitons in gate-dependent photoluminescence (PL) measurements can be used as a proxy for carrier density in 2D materials \cite{Mak2012,Mouri2013,Ross2013}. 
In \cref{fig:2}a, we plot the PL map of our trilayer CrSBr device as a function of gate voltage ($\Delta V = 0$). With increasing $V_{\text{b}} = V_{\text{t}}$, the intensity of the trion emission at $\SI{1.34}{\eV}$ suddenly increases at $V_{\text{b}} = V_{\text{t}} \approx \SI{-0.8}{\volt}$, concurrently with a dimming of the excitonic peak at $\SI{1.37}{\eV}$.
This is consistent with the neutral excitons binding to additional electrons, and 
suggests an intrinsic n-doping of our sample (as the trions dominate the spectrum at $V_{\text{b}} = V_{\text{t}} = \SI{0}{\volt}$). 
Therefore, we conclude that below $V_{\text{b}} = V_{\text{t}} \approx \SI{-0.8}{\volt}$, the Fermi level lies in the bandgap and the sample is undoped. \\
In \cref{fig:2}b, the gray data points show an exemplary PL spectrum of our trilayer CrSBr device in this undoped regime without applied field, i.e. $n = 0$ and $\Delta V = \SI{0}{\volt}$. 
We observe strong emission at $X_{\mathrm{B}} \approx \SI{1.374}{eV}$ with a shoulder at lower energies stemming from $X_{\mathrm{B}}' \approx \SI{1.370}{\eV}$.
In previous reports, $X_{\mathrm{B}}$ and $X_{\mathrm{B}}'$ were assigned to the excitonic transition between top valence and second lowest conduction band \cite{Wilson2021,Klein2023,Farsane2023,Lin2024_2}. 
Their energetic splitting in samples with > 2 layers has been attributed to different dielectric screening in the outer (B) versus inner (B') layers, leading to a layer-resolved response of thin samples to electrostatic gating \cite{Farsane2023}. 
Around $\SI{1.34}{\eV}$, we find an amalgamation of several peaks even in the absence of trions (as $n = 0$), which can be assigned to the $X_A$ exciton (transition between top valence and bottom conduction band) and phonon replicas \cite{Wilson2021,Klein2023,Farsane2023,Lin2024_2}. We note that a different interpretation of the emission at 1.34 and $\SI{1.37}{eV}$ has recently been given, in which the latter has been attributed to surface and the former to bulk excitons \cite{Shao2025}. However, the thickness-dependent relative intensities of $X_{\mathrm{B}}$ and $X_{\mathrm{B}}'$ observed in our experiments support their assignment as excitons in the outer and inner layers (gate-dependent PL of thicker samples in SI section 3).\\
For sufficiently high electron doping ($n > 0$), we see emission from trions at $X_{\mathrm{B}}^- \approx \SI{1.348}{\eV} $ and $X_{\mathrm{B}}'^- \approx \SI{1.344}{\eV}$ (doped spectra in \cref{fig:2}b, fitted exciton and trion peaks in upper panel).
We note that the binding energies of the outer and inner layer trions $X_{\mathrm{B}}^--X_{\mathrm{B}} = X_{\mathrm{B}}'^--X_{\mathrm{B}}' = \SI{26}{\milli\eV}$ match, confirming the assignment of these peaks. Therefore, we use the relative intensities $I_{\mathrm{B}} = \frac{I_{X_{\mathrm{B}}}}{I_{X_{\mathrm{B}}^-}}$ and $I_{\mathrm{B}'} = \frac{I_{X_{\mathrm{B}}'}}{I_{X_{\mathrm{B}}'^-}}$ as indicators for the carrier densities of the outer layers, $n_t+n_b$, and the middle layer, $n_m$, respectively. Field-dependent measurements confirm this assignment:  In \cref{fig:2}b, we show two selected PL spectra for opposite fields $F_{\mathrm{z}} \propto \Delta V$, but equivalent doping level $n$ (orange: $\Delta V < 0$, purple: $\Delta V > 0$). While the spectral weight $I_{\mathrm{B}}$ is nearly identical for the two (as either top or bottom layer become doped), $I_{\mathrm{B}'}$ differs for $\Delta V > 0$ and $\Delta V < 0$ (see fits in upper panel of \cref{fig:2}b).
\cref{fig:2}c shows the complete dependence of $I_{\mathrm{B}'}$ on $\Delta V$, in which the points at $\Delta V$ and $- \Delta V$ have equivalent doping (complete gate-dependence of $I_{\mathrm{B}}$, $I_{\mathrm{B}'}$ in SI section 3). Generally, the intensity $I_{\mathrm{B}'}$ is smaller for positive than negative polarity of $\Delta V$, which suggests a built-in electrical field $F_{\mathrm{z},0}$, possibly originating from surface charges \cite{Torres2023}. 
\\
We link $I_{\mathrm{B}}$ and $I_{\mathrm{B}'}$ to the layer-resolved electron densities $n_i$ and electric fields $F_{\mathrm{z},ij}$ (where $i,j$ = top, middle and bottom layers) using a simple electrostatic model \cite{Pisoni2019,Bolotin2023}. 
The main free parameters are the built-in field $F_{\mathrm{z},0}$ and the carrier density $n_0$ at $V_{\text{b}} = V_{\text{t}} = \SI{0}{\volt}$. The former leads to an offset between the conduction bands between the layers, and the latter defines the initial position of the chemical potential.
For the behavior of the layer-resolved carrier densities to resemble the experimentally obtained PL data we arrive at $n_0 \approx \SI{2.2e12}{\per\centi\metre\squared}$ and $F_{\mathrm{z},0} \approx -\SI{0.01}{\volt\per\nano\metre}$ (details of the model and gating maps of $n_i$ and $F_{\mathrm{z},ij}$ in SI section 3). Comparing the resulting $n_i$ as a function of $V_{\text{b}} = V_{\text{t}}$ (\cref{fig:2}d, lower panel) to the ratios $I_{\mathrm{B}}$ and $I_{\mathrm{B}'}$ extracted from \cref{fig:2}a (\cref{fig:2}d, upper panel), we find a good alignment of the drop in both exciton-to-trion ratios around $\sim -\SI{0.8}{\volt}$ with the onset of doping between $\sim -0.5$ and $- \SI{0.2}{\volt}$. 
The doping-dependence of the total carrier density of all three layers $n = \sum_i n_i$ (\cref{fig:2}d, gray dashed line in lower panel) follows that of the magnon frequencies in \cref{fig:1}c with a small offset. 
This suggests that carrier-density induced modifications of magnetic properties contribute to the changes in magnon dynamics. However, the asymmetry with $\Delta V$ shown in the full gating maps of $f_{\text{IP}}$ and $f_{\text{OP}}$ in \cref{fig:1}d, e cannot be explained by changes in $n$ alone, indicating a dependence on the electric fields $F_{\mathrm{z},ij}$, as well. \\
To explain how a doping-dependent layer asymmetry influences the magnonic behavior, we use a layer-resolved macrospin model. This approximation assumes strongly coupled spins within individual layers, such that their collective motion can be described by a single macrospin $\vec{m}_i$. Due to the long magnon lifetime ($> \SI{900}{\pico\second}$) observed in \cref{fig:1}b, and because we are mainly interested in the magnon frequencies, we omit damping effects. The dynamics of each macrospin are then governed by the layer-dependent effective field $\vec{H}_{\mathrm{eff},i}$ in the Landau-Lifshitz (LL) equation:
\begin{equation}
    \frac{\mathrm{d}\vec{m}_{i}}{\mathrm{dt}} = -\gamma_i \vec{m}_{i}\times\vec{H}_{\mathrm{eff},i}.
    \label{eq:LLG}
\end{equation}
Here, $\gamma_i = \frac{g\mu_{i}}{\hbar}$ is the layer-dependent gyromagnetic ratio, physically representing the overall layer magnetization. The effective field is given by
\begin{equation}
    \vec{H}_{\mathrm{eff}_{i}} = -\nabla_{\vec{m}_{i}} E = \vec{H}_{0} - \sum_{j} H_{\mathrm{E}_{ij}} \vec{m}_j + H_{\mathrm{a}_i}m_i^{\mathrm{a}}\hat{a} + H_bm_i^{\mathrm{b}}\hat{b},
    \label{eq:efffield}
\end{equation} 
with the external field $\vec{H}_0$, the interlayer exchange coupling $H_{\mathrm{E}_{ij}}$ between layers $i$ and $j$, and two anisotropies $H_{\mathrm{b}}$ and $H_{\mathrm{a}_i}$ along the easy $\hat{b}$- and intermediate $\hat{a}$-axis. The macrospin components $m_i^{\mathrm{a}}$ and $m_i^{\mathrm{b}}$ point in those same directions. In contrast to conventional models, the gyromagnetic ratio, interlayer exchange and intermediate axis anisotropy are taken to be explicitly layer-dependent.
To solve \cref{eq:LLG}, the external field of $H_{0} \approx \SI{0.1}{\tesla}$ is taken to be purely in the out-of-plane direction. The excursions $\delta \vec{m}_i$ of the macrospins away from the equilibrium $\vec{m}_{i,0}$ are assumed small, allowing linearization of the LL equation. We then extract the magnon frequencies numerically (details in SI section 5).\\  
To ensure the validity of the layer-resolved model, it is instructive to compare the numerical dispersion in the case of a bulk-like (100 layer) sample to the analytical solution routinely used to model the bulk case (details in SI section 4). 
In \cref{fig:3}a, we show that the numerical model gives a low-frequency acoustic (blue points) and high-frequency optical (red points) magnonic branch consisting of bulk modes, in good agreement with the analytical result (solid lines). 
The acoustic and optical modes at $k_z = 0$ correspond to the $f_{\text{IP}}$ and $f_{\text{OP}}$ excited and detected in previous pump-probe experiments \cite{Bae2022}. 
Remarkably, two additional degenerate modes with low frequency emerge in the numerical model (marked with a star in \cref{fig:3}a). 
These are edge modes with imaginary $k_z$ which decay exponentially into the bulk sample and are a result of the open boundary conditions used. \\
In thin samples, the magnon modes captured by the layer-resolved model begin to deviate from those seen in the bulk. \cref{fig:3}b shows the magnon frequencies as a function of device thickness (expressed in layer number $N$). 
For $N > 10$, the eigenfrequencies of the bulk modes (between $\sim 20 -\SI{28}{\giga\hertz}$) as well as the edge modes (below $\sim \SI{20}{\giga\hertz}$) remain nearly  thickness-independent, in agreement with previous experimental reports \cite{Bae2022}. 
When reducing the layer number to $N < 10$, the maximum and minimum frequencies of the bulk modes become thickness-dependent due to the stronger influence of the edge layers, which experience a smaller exchange coupling. 
Additionally, the edge modes become non-degenerate, and they spread across the whole sample instead of being localized at one side. 
In trilayers, we find that only the highest and lowest frequency magnon couple efficiently to a photothermal excitation, so we consider these modes as $f_{\text{OP}}$ and $f_{\text{IP}}$ in our experiments (extended discussion in SI section 6).\\
Having established the layer-resolved magnetic properties, we now turn to explaining the $V_{\text{b}}$-$V_{\text{t}}$-dependent behavior of $f_{\text{IP}}$ and $f_{\text{OP}}$ (\cref{fig:1}d, e). 
In general, various physical processes can result in carrier-density or electric-field-dependencies of the parameters of \cref{eq:LLG,eq:efffield}. 
To determine compatible mechanisms, we analyze the dependence of the magnon modes on overall changes in $\gamma (= \gamma_{\text{t}} = \gamma_{\text{m}} = \gamma_{\text{b}})$, $H_{\text{a}} (= H_{\text{a}_{\text{t}}} = H_{\text{a}_{\text{m}}} = H_{\text{a}_{\text{b}}})$ and $H_{\text{E}} (= H_{\text{E}_{\text{tm}}} = H_{\text{E}_{\text{mb}}})$ in the numerical model within realistic ranges, see \cref{fig:3}c-e. 
We observe that both $f_{\text{IP}}$ and $f_{\text{OP}}$ increase linearly with $\gamma$ at similar rates (\cref{fig:3}c). 
Conversely, an increasing $H_{\mathrm{a}}$ leads to a decrease in both $f_{\text{IP}}$ and $f_{\text{OP}}$, with a weaker effect on the latter (\cref{fig:3}d). 
Both frequencies also increase with $H_{\text{E}}$ at different rates, with a stronger influence on $f_{\text{OP}}$ (\cref{fig:3}e). We only include one of the two anisotropies to be layer-dependent, as increasing $H_{\mathrm{a}}$ has a similar influence on the frequencies as decreasing $H_{\mathrm{b}}$ (SI section 5). We therefore fix $H_{\mathrm{b}} = \SI{1.3}{\tesla}$ close to previously reported values in the simulations \cite{Bae2022}.\\
From these considerations, we find that the experimentally observed shift of $f_{\text{IP}}$ and $f_{\text{OP}}$ at the same rate with respect to carrier density in \cref{fig:1}c could be explained by a dependence of $\gamma_i$ on $n_i$. 
In contrast, the asymmetric shifts of $f_{\text{IP}}$ and $f_{\text{OP}}$ vs. applied field in \cref{fig:1}d, e suggest a competing decrease of $H_{\mathrm{a}_i}$ and increase of $H_{\mathrm{E}_{ij}}$ as a function of $F_{\mathrm{z},ij}$. 
\\
We therefore suggest the following minimal set of dependencies:
\begin{equation}
    H_{\mathrm{E}_{ij}} = H_{E_0} + \nu_{\mathrm{E}} F_{\mathrm{z},ij} 
    \label{eq:hedep}
\end{equation}
\begin{equation}
    H_{\mathrm{a}_i} = H_{\mathrm{a}_0} + \nu_{\mathrm{a}} \sum_{\left< i,j \right>}F_{\mathrm{z},ij}
    \label{eq:hadep}
\end{equation}
\begin{equation}
    \gamma_{i} = \gamma_0 + \eta_{\mathrm{\gamma}} n_i 
    \label{eq:gammadep}
\end{equation}
Using \cref{eq:hedep,eq:hadep,eq:gammadep}, we fit the frequency gating maps in \cref{fig:1}d, e to our layer-dependent LL model using the $n_i$ and $F_{\mathrm{z},ij}$ gating maps from the electrostatic model in SI Fig. 5.
The fitted frequency maps shown in \cref{fig:1}f, g reproduce all basic trends of the data, in particular the opposing $V_{\text{t}}$- and $V_{\text{b}}$-sensitivity of the two magnon modes.
In the doping-dependency of the modes in \cref{fig:1}c, the solid lines also depict the model results and show good agreement with the data, apart from a shifted onset of frequency changes. 
The resulting fitting parameters are summarized in the first row of \cref{tab:fittrilayer} (errors for the trilayer estimated by varying the fixed variable $\gamma_0 = \SI{155(15)}{\giga\hertz\per\tesla}$ close to previously reported values \cite{Bae2022}). 
\\
Several doping- and electric-field-related effects can explain the physical origin of the influence of $n_i$ and $F_{\mathrm{z},{ij}}$ on magnetic properties.  First, the position of the chemical potential -- controlled by  $n$ -- affects the imbalance between the occupied spin-up and spin-down states, affecting the magnetic moment $\mu$ and in turn the gyromagnetic ratio $\gamma$ of the corresponding macrospin \cite{Han2024}. In addition, an $n$-dependent increase of the Coulomb repulsion influences the antiferromagnetic exchange constant \cite{Farsane2023,Kugel1973,Kugel1982}.
Second, the relative population of states derived from $e_g$ and $t_{2g}$ orbitals in one layer is affected by both chemical potential and external field.  This leads to a corresponding dependence of both exchange \cite{Kugel1973,Kugel1982,Mazurenko2006} and anisotropy terms \cite{Ziebel2024,Maruyama2009,He2011,Rana2019}. Third, a perpendicular electric field shifts the energy bands associated with different orbitals in neighboring layers with respect to each other, again influencing the exchange coupling parameter \cite{Wang2023_2} (SI section 14). \\
We believe that the suggested dependencies in \cref{eq:hedep,eq:hadep,eq:gammadep} are the governing mechanisms of the gate dependency of magnons in CrSBr. The inclusion of additional $n_i$-dependent terms in the Hamiltonian, e.g. $H_{\mathrm{E}_{ij}}$ and $H_{\mathrm{a}_i}$, has not significantly improved the fits.
For the trilayer sample, we predict a change in $H_{\mathrm{E}_{ij}}$ between neighboring layers of up to $\sim\SI{100}{\milli\tesla}$ as a result of the applied electric field, while the tunability of in $H_{\mathrm{a}_i}$ reaches almost $\sim\SI{200}{\milli\tesla}$. These changes of effective fields are an order of magnitude stronger than the previously reported gate-tunable internal field in a \SI{10}{\nano\metre} thick \ce{Cr2Ge2Te6} sample \cite{Hendriks2024} and surpass the voltage-controlled magnetic anisotropy induced, for example, in CoFeB samples \cite{Rana2019,Li2017,Nozaki2013}. As $\gamma_i \propto n_i$, the gyromagnetic ratio changes most significantly in the top layer due to its strong doping, increasing from $\gamma_{\text{t}} = 155$ to  $\sim\SI{180}{\giga\hertz\per\tesla}$, which is larger than the predicted increase of magnetic moment for our doping levels in a similar system, CrSeBr \cite{Han2024} (complete gating maps of $\gamma_i$, $H_{\mathrm{a}_i}$, $H_{\mathrm{E}_{ij}}$ and discussion in SI section 7).
\\
Additionally, the $n_i$- and $F_{\mathrm{z},{ij}}$-dependencies found for the trilayer predict the gate-dependent magnon frequency changes in thicker samples. The $V_{\text{b}}$-dependence of $f_{\text{IP}}$ and $f_{\text{OP}}$ in a 5- and 8-layer device, shown in comparison with the previous trilayer data, is plotted in \cref{fig:4}a and b (additional 8-layer data in the SI section 8). 
The magnitude of the gating effects decrease with thickness: In the 5-layer device, the maximum shift of $f_{\text{OP}}$ is only $\SI{0.5}{\giga\hertz}$, while the 8-layer device shows no discernible frequency shift. 
The 5- and 8-layer device also reproduce the insensitivity of $f_{\text{IP}}$ to $V_{\text{b}}$ observed in the trilayer. 
Using $\nu_a$, $\nu_E$ and $\eta_{\mathrm{\gamma}}$ extracted from the trilayer data, we model the expected frequency shifts in the thicker devices. The fitted $H_{\mathrm{a}_0}$ and $H_{\mathrm{E}_0}$ differ from the trilayer results, possibly due to different prestrain across devices (values and estimated errors in \cref{tab:fittrilayer}). 
The modeled gate-dependent changes (solid lines in \cref{fig:4}a, b) agree with our experimental observations of reduced gate-sensitivity with higher layer number. 
In thicker devices, both carrier density and fields are induced mainly around the outer layers, with the inner ones being screened from the gates (electrostatic modeling for 5- and 8-layer device in SI section 3). 
The influence of these outer layers on the magnon frequencies decreases with increasing layer number $N$, illustrated in \cref{fig:4}c, d. The magnon frequency shifts due to changes of the anisotropy (\cref{fig:4}c) or exchange interaction (\cref{fig:4}d) for one outer layer decrease drastically going from three to ten layers in the layer-resolved model.\\ 

%% file: discussion.tex
\section*{Discussion}
In summary, we have demonstrated the gate-tunability of the in- and out-of-phase magnon modes in trilayer CrSBr by up to 10 \% due to unprecedentedly large changes in interlayer exchange and anisotropy fields. By using the contrasting response of the two magnons to top and bottom gates, their separate control is possible. We explain this gate-dependence in a Landau-Lifshitz model with layer-dependent gyromagnetic ratio, interlayer exchange and anisotropy. 
By coupling the interlayer exchange and intermediate-axis-anisotropy to the electric fields and the gyromagnetic ratio to the electron densities across layers, we arrive at a minimal model to fit our data. 
The model explains most trends in the data including the contrasting response of the two magnon modes to top and bottom gates, their doping response, as well as the observed thickness dependence. 
In particular, the gate-tunability of magnons in CrSBr drops to $\approx \SI{0.5}{\giga\hertz}$ for a 5-layer device and vanishes for 8 layers. 
In future experiments, this tunability could be further expanded by new gating techniques which surpass electrostatic gating in terms of the achievable doping and field \cite{Weintrub2022,Domaretskiy2022,Bolotin2023}. 
Our findings open up new possibilities for using the on-chip control of magnons for magnonic circuits, e.g., for a phase shifter used in logic gates. By tuning magnons in- and out of phase with each other, they can interfere destructively (bit = 0) or constructively (bit = 1), thus transmitting information \cite{Chumak2015,Rana2018_2}. \\

%% file: main.tex

\begin{table*}[h]

    \centering
    \begin{tabular}{|c|c|c|c|c|c|}
        \hline
        & $H_{\mathrm{E}_0}$ (T) & $H_{\mathrm{a}_0}$ (T) & $\nu_{\mathrm{a}}$ ($\si{\tesla\nano\metre\per\volt}$)& $\nu_{\mathrm{E}}$ ($\si{\tesla\nano\metre\per\volt}$)& $\eta_{\gamma}$ ($\si{\giga\hertz\centi\metre\squared\per\tesla}$)\\ 
        \hline
        \hline
        trilayer & $0.20 \pm 0.05$ & $0.97 \pm 0.05 $ & $-0.8 \pm 0.2$ & $-1.0 \pm 0.2$ & $2.3 \pm 0.3$ \\
        \hline
        5-layer & $0.28 \pm 0.02$ & $0.99 \pm 0.01$ & $-0.8 \pm 0.3$ & $-1.0 \pm 0.7$ & $2.3 \pm 0.5$  \\
        \hline
        8-layer & $0.25 \pm 0.01$ & $0.91 \pm 0.02 $& $-0.8 \pm 0.4$ & $-1.0 \pm 0.4$ & $2.3 \pm 1.8$  \\
        \hline
    \end{tabular}
    \caption{First row shows the fitting results of the trilayer sample $V_b$-$V_t$-maps in \cref{fig:1}g, h using the layer-resolved macrospin model. The easy-axis anisotropy $H_{\mathrm{b}} = \SI{1.3}{\tesla}$ and initial gyromagnetic ratio $\gamma_0 = \SI{155}{\giga\hertz\per\tesla}$ are fixed. The errors for the trilayer modeling were estimated by varying $\gamma_0$ from 140 to $\SI{170}{\giga\hertz\per\tesla}$. For the 5- and 8-layer device, the experimental frequencies were fitted to the macrospin model with $\nu_{\mathrm{a}}$, $\nu_{\mathrm{E}}$ and $\eta_{\gamma}$ fixed to the trilayer values. The resulting $H_{\mathrm{a}_0}$ and $H_{\mathrm{E}_0}$ differ slightly between samples. The errors of the 5- and 8-layer results are estimated by fitting the experimental data while only fixing $H_{\mathrm{b}}$ and $\gamma_0$ (as in the trilayer case).}
    \label{tab:fittrilayer}
\end{table*}

%% file: methods.tex
\section*{Methods}
\subsubsection*{Sample fabrication and crystal growth}
Samples were fabricated using a dry-transfer technique after exfoliation on PDMS. The contacts were patterned by electron beam lithography followed by evaporation of 3 nm Cr and 80 nm Au.\\
The crystals of CrSBr were synthesized by chemical vapor transport and subsequently characterized by X-ray diffraction, crystal diffraction (powder and single crystal), transmission electron microscopy, 
energy dispersive X-ray analysis, Raman and IR spectroscopy, superconducting quantum interference device magnetometry and magneto-transport measurements, as reported in \cite{Samuel}.

\subsubsection*{Tr-reflectivity measurements}
All tr-reflectivity measurements have been performed in a single-color pump and probe scheme. The laser source is a wavelength-tunable Ti:sapphire laser (Coherent Chameleon Ultra II, $\tau_{pulse} \approx \SI{140}{\pico\second}, f_{rep} = \SI{80}{\mega\hertz}$).
We spatially separate pump and probe beam before entering a reflective objective (Thorlabs LMM40x-P01) and focusing on the sample, into spot sizes of $d_{\text{probe}} \approx \SI{1}{\micro\metre}, d_{\text{pump}} \approx \SI{3}{\micro\metre}$. 
Thanks to the spatial separation of the reflected beams we then filter out the pump beam using an iris. 
The probe light is measured with a home-built photodetector (Hamamatsu photodiodes) using a lock-in amplifier synced to a chopper in the pump beam ($f = \SI{2.1}{\kilo\hertz}$). More details are provided in SI section 1.

\subsubsection*{PL measurements}
All PL measurements were conducted using a continuous wave laser with an excitation wavelength of $\SI{670}{\nano\metre}$ and a laser power of $\SI{25}{\micro\watt}$. The incoming laser light was linearly polarized along the $\hat{b}$-axis of the CrSBr flake. A refractive objective (Olympus LMPlanFL N 50x/0.50) focused the spot to a size of $\approx \SI{1}{\micro\meter}$. The resulting copolarized PL was captured with the Kymera 193i Spectrograph. 

\subsubsection*{Dielectric constant of CrSBr}
The dielectric constant of CrSBr used in the electrostatic model was calculated using Hubbard corrected DFT energy functionals (LDA+U). The complete dielectric function is shown in SI Fig. 4.

\section*{Acknowledgements}
We acknowledge the German Research Foundation (DFG) for financial support through the Collaborative Research Center TRR 227 Ultrafast Spin Dynamics projects B08, B03 and A04, as well as the priority programme SPP2244. We thank T. Kampfrath and G. Woltersdorf for useful discussions. We thank B. Höfer for technical support.

\section*{Contributions}
N.S. and K.I.B. conceived the project. A.M.K. and D.Y. designed the experimental setup. N.S. and S.K. prepared the samples. N.S. performed the measurements. S.K. provided the original code for electrostatic modeling. N.S., L.S., K.I.B., and P.B. developed the layer-resolved macrospin model. N.S. performed the electrostatic and macrospin modeling of the results. N.S. analyzed the data. S.M.V. and E.C. grew the CrSBr crystals. T.T. and K.W. grew the hBN crystals. D.G. and S.S. performed the DFT calculations of the dielectric function. K.I.B., C.K.S. and S.E. supervised the project. N.S. and K.I.B. wrote the manuscript with input from all co-authors.

%% file: figures.tex
\begin{figure*}
    \centering
    \includegraphics[width=\linewidth, trim = 0mm 20mm 0mm 20mm,clip=true]{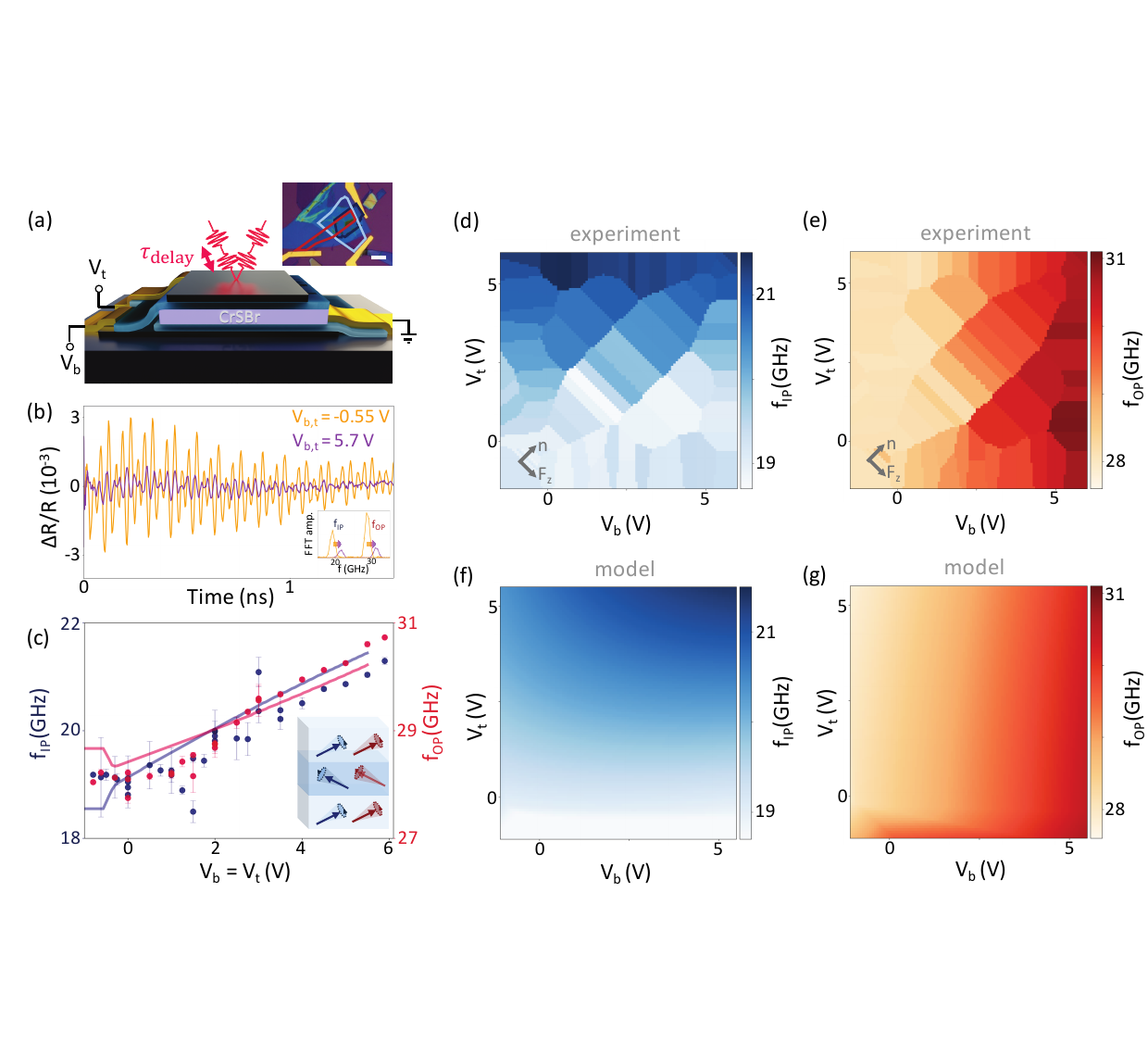}
    \caption{\textbf{Magnons in double-gated trilayer CrSBr}
    (a) Device scheme of a trilayer CrSBr flake (purple) encapsulated in hBN (light blue) with a top and bottom graphite gate (gray). 
    The magnons are excited with an ultrafast pump laser pulse, and read out by a probe pulse after $\tau_{\text{delay}}$. 
    Inset shows optical image of the trilayer device (red: CrSBr flake, blue: top and bottom graphite), the scale bar corresponds to $\SI{10}{\micro\metre}$. 
    (b) Two examples for time-resolved reflectivity traces after background subtraction, for low ($V_{\text{b}} = V_{\text{t}} = \SI{-0.55}{\volt}$, orange) and high ($V_{\text{b}} = V_{\text{t}} = \SI{5.7}{\volt}$, violet) electron doping. 
    The drop in signal intensity stems from the doping-dependent changes in the tr-reflectivity spectra (see SI Fig. 1). An in-phase mode $f_{\text{IP}}$ and out-of-phase mode $f_{\text{OP}}$ are extracted from the traces using Fast Fourier Transform, shown in the inset. Electron doping shifts the magnon frequencies from 19.1 to $\SI{21.1}{\giga\hertz}$ and 28.1 to $\SI{31}{\giga\hertz}$, respectively.
    (c) The complete doping response ($V_{\text{b}}  = V_{\text{t}}$) of $f_{\text{IP}}$ (blue, see inset) and $f_{\text{OP}}$ (red, see inset) shows that the frequency increase only starts after a threshold voltage. 
    The solid lines are fits using the layer-resolved macrospin model.
    (d-e) The complete $V_{\text{b}}$-$V_{\text{t}}$-dependences of (d) $f_{\text{IP}}$  and (e) $f_{\text{OP}}$ plotted as Voronoi diagrams show a complex response which is opposite with electric field (arrows mark doping $n$ and electric field $F_{\mathrm{z}}$ direction). 
    (f-g) Using a layer-dependent macrospin model connected to an electrostatic model we can reproduce the experimental trends in the simulated $V_{\text{b}}$-$V_{\text{t}}$-diagrams for both (f) $f_{\text{IP}}$ and (g) $f_{\text{OP}}$.
    }
    \label{fig:1}
\end{figure*}                                                                               
\begin{figure*}[h]
    \centering
    \includegraphics[width=\linewidth, trim = 0mm 50mm 0mm 50mm,clip=true]{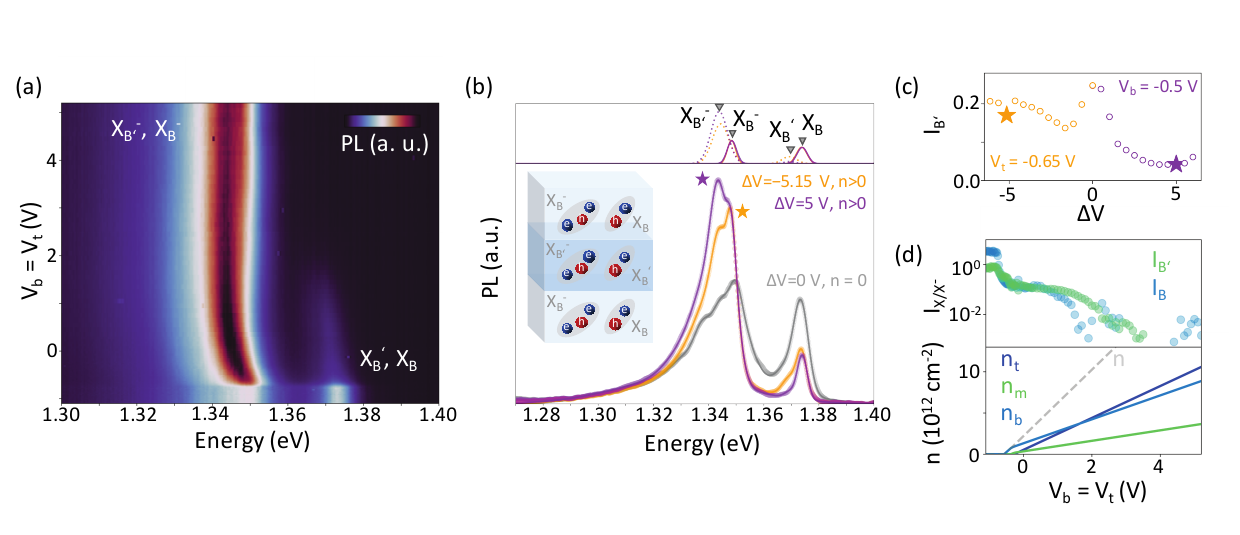}
    \caption{\textbf{Dual-gate-dependence of photoluminescence in trilayer CrSBr}
    (a) Doping-dependent PL map of the trilayer CrSBr showing the dimming of neutral exitons $X_{\mathrm{B}}$ and $X_{\mathrm{B}}'$ around $\sim\SI{1.37}{\eV}$ and brightening of the trions $X_{\mathrm{B}}^-$ and $X_{\mathrm{B}}'^-$ at $\sim\SI{1.35}{\eV}$ with increased electron density. 
    (b) Exemplary PL spectra for undoped (gray) and doped (orange, purple) trilayer CrSBr (dots are datapoints, solid lines are fits using a sum of gaussians).
    The upper panel shows the fitted exciton and trion peaks for the doped spectra -- both the exciton and trion peaks split in two due to them residing in the outer (B, solid lines) vs. the middle layer (B', dashed lines), as shown in the inset sketch.
    Under the same overall electron density $n$ in the sample, but opposite electric field polarity $\propto \Delta V$ (orange: $\Delta V > 0$, purple: $\Delta V < 0$) the exciton-to-trion ratio $I_{\text{B}}$ is almost identical (compare orange and purple solid lines in upper panel), while $I_{\text{B'}}$ is sensitive to the field direction (compare orange and purple dashed lines in upper panel). 
    (c) Extracted $I_{\text{B'}}$ for fixed $V_{\text{t}} = \SI{-0.65}{\volt}$ (orange) or fixed $V_{\text{b}} = \SI{-0.5}{\volt}$ (purple), but varying $\Delta V = V_{\text{t}} - V_{\text{b}}$ (spectra from (b) marked with stars). For the same doping level (graph is symmetric with doping), the ratio is lower for a positive $\Delta V$, due to a higher electron density $n_{\text{m}}$ for positive field direction. 
    (d) Extracted ratios $I_{\text{B}}$ and $I_{\text{B'}}$ from (a) in the top panel, and layer-resolved carrier densities from electrostatic modeling in the lower panel. The drop of the ratios coincides with the modeled increase of $n$ (gray).
    }
    \label{fig:2}
\end{figure*}
\begin{figure*}[h]
    \centering
    \includegraphics[width=\linewidth, trim = 0mm 30mm 0mm 30mm,clip=true]{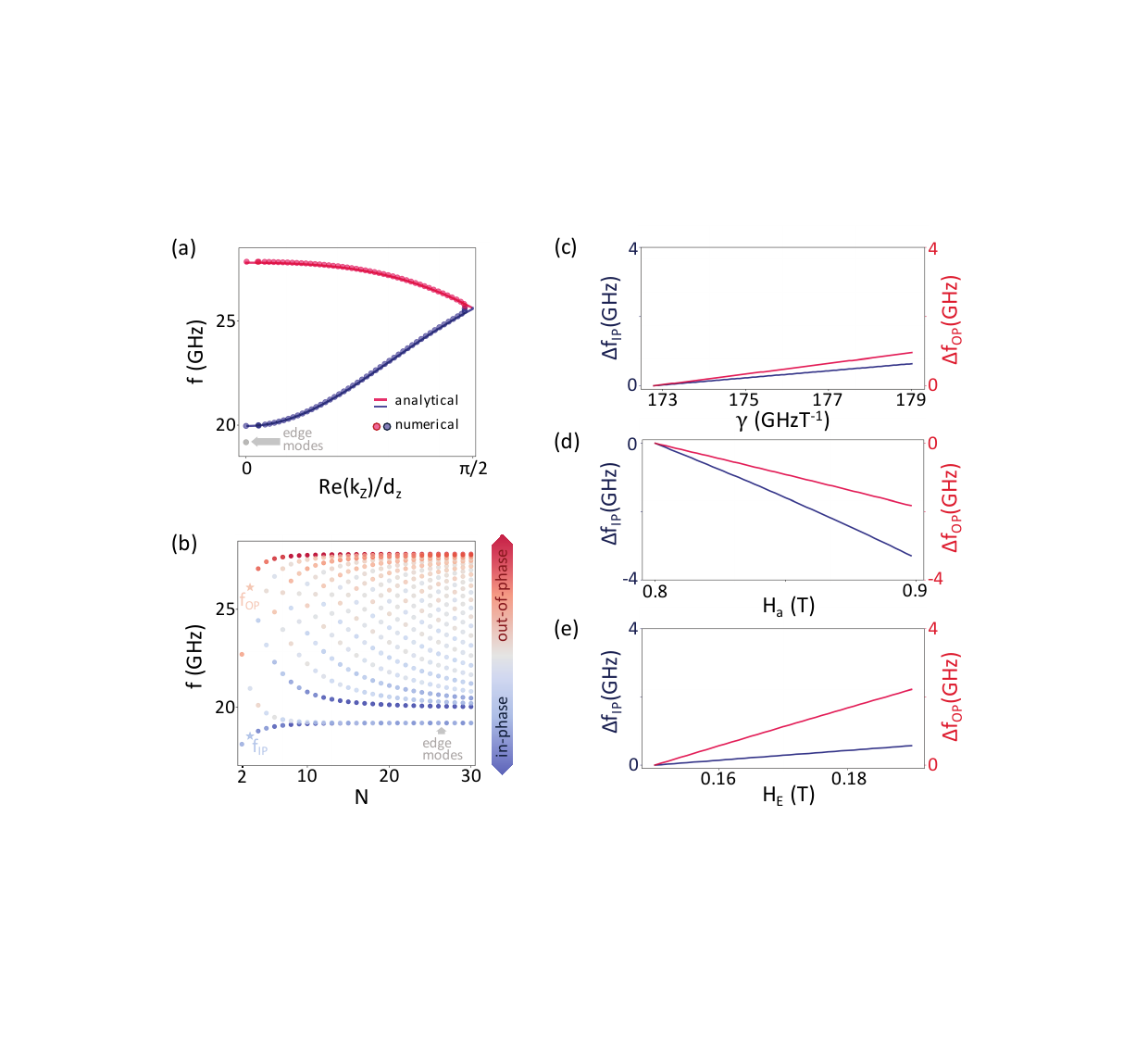}
    \caption{\textbf{Layer-resolved macrospin model}
    (a) Comparison of the magnon dispersion of $k_{\mathrm{z}}$ ($k_{\mathrm{x}} = k_{\mathrm{y}} = 0$; $d_{\mathrm{z}}$ is interlayer spacing) calculated with the analytical bulk model (solid lines) and the layer-resolved numerical model for 100 layers (dots). Both models show an acoustic (blue) and an optical (red) magnon branch. The edge modes (gray) in the layer-resolved model are a result of the open boundary conditions.
    (b) Numerical magnon frequencies as a function of layer number $N$ calculated with the layer-resolved model. The color-coding shows the coupling of the modes to in- or out-of-phase excitations (details in SI section 6). For high $N$, the degenerate edge modes lie $\approx \SI{1}{\giga\hertz}$ below the bulk modes of the acoustic branch. For $N < 10$, the edge modes split. For $N = 3$, we identify the lowest and highest eigenvalue as the experimental $f_{\text{IP}}$ and $f_{\text{OP}}$, respectively (marked by stars).
    (c-e) Influence of changing the (c) gyromagnetic ratio $\gamma$, (d) anisotropy $H_{\text{a}}$ or (e) interlayer exchange $H_{\text{E}}$ in all layers in the layer-resolved model. Both frequencies shift with $\gamma$ at approximately the same rate, while their sensitivity to $H_{\mathrm{a}}$ and $H_{\mathrm{E}}$ differs. For all plots, the respective fixed parameters are $\gamma = \SI{176}{\giga\hertz\per\tesla}$, $H_{\text{b}} = \SI{1.3}{\tesla}$, $H_{\text{a}} = \SI{0.9}{\tesla}$ and $H_{\text{E}} = \SI{0.15}{\tesla}$ ($\SI{0.3}{\tesla}$ in the analytical bulk model in (a)). 
    }
    \label{fig:3}
\end{figure*}
\begin{figure*}[h]
    \centering
    \includegraphics[width=\linewidth, trim = 0mm 50mm 0mm 50mm,clip=true]{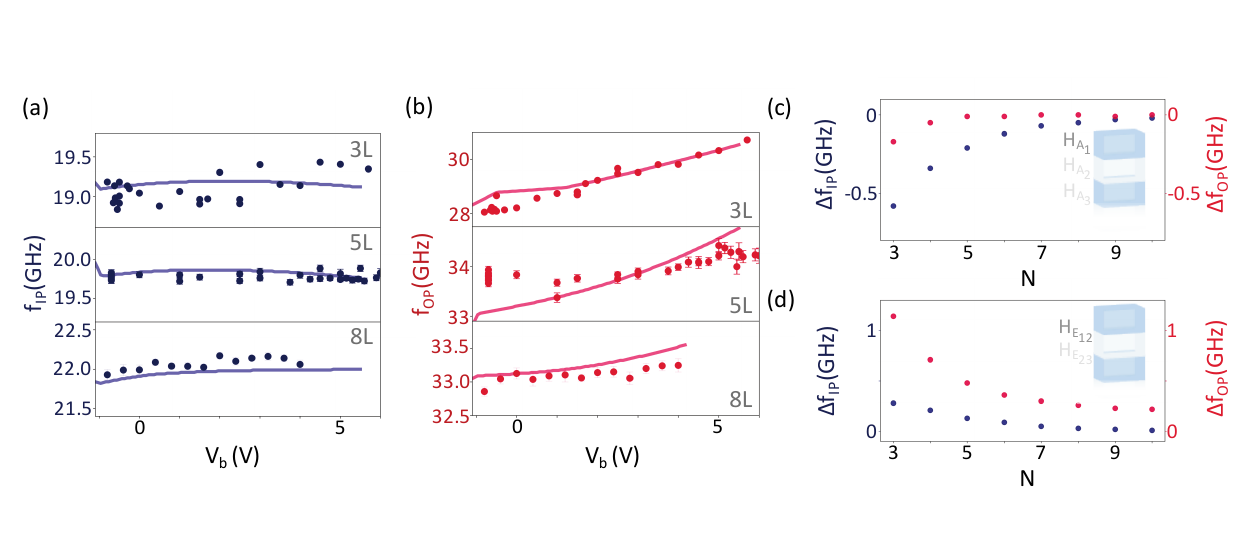}
    \caption{\textbf{Thickness-dependence of magnon tunability}
    (a-b) Comparison of the $V_{\text{b}}$-dependency of (a) $f_{\text{IP}}$ and (b) $f_{\text{OP}}$ for three, five and eight layers of CrSBr. The solid lines show the frequencies fitted with the macrospin model (fitting parameters in \cref{tab:fittrilayer}).
    In (a), $f_{\text{IP}}$ is insensitive to the bottom gate across all devices, while the tunability of $f_{\text{OP}}$ in (b) drops with layer number. Note the difference in frequency scale in (b) for better visibility of the small shift in the thick devices.
    (c) Modeled shifts $\Delta f_{\text{IP}}$ and $\Delta f_{\text{OP}}$ when changing $H_{\text{a}_1}$ from 0.8 to $\SI{0.9}{\tesla}$ ($H_{\mathrm{a}_i}$ of all other layers fixed to $\SI{0.9}{\tesla}$) as a function of layer number $N$. The shift drops to almost 0 for both modes in 10 layers. 
    (d) Varying $H_{E_{12}}$ from 0.15 to $\SI{0.19}{\tesla}$ (other exchange couplings fixed to $\SI{0.15}{\tesla}$) also results in smaller $\Delta f_{\text{IP}}$ and $\Delta f_{\text{OP}}$ for higher $N$.
    }
    \label{fig:4}
\end{figure*}

%% file: main_si.tex
\begin{titlepage}
  \centering
  \vskip 500pt
  \LARGE Supplementary information: Tunable magnons in a dual-gated 2D antiferromagnet \par
  \vskip 3em
  \large Nele Stetzuhn$^{1,2}$, Abhijeet M. Kumar$^{1}$, Sviatoslav Kovalchuk$^{1}$, Denis Yagodkin$^{1}$,\\ Louis Simon$^{1}$, Samuel Mañas-Valero$^{3,4}$, Eugenio Coronado$^{3}$, Takashi Taniguchi$^{6}$, \\Kenji Watanabe$^{6}$, Deepika Gill$^{2}$, Sangeeta Sharma$^{1,2}$, Piet Brouwer$^{1}$,\\ Clemens von Korff Schmising$^{2,5}$, Stefan Eisebitt$^{2,5}$, Kirill I. Bolotin$^{1}$\\
  \vskip 1.5em
    \small
    $^1$ Department of Physics, Freie Universität Berlin, Arnimallee 14, 14195 Berlin, Germany\\%
    $^2$ Max Born Institute for Nonlinar Optics and Short-Pulse Spectroscopy, Max-Born-Str. 2A, 12489 Berlin, Germany\\
    $^3$ Instituto de Ciencia Molecular, Universidad de Valencia, Dr. Moliner 50, Burjasot, 46100, Spain\\
    $^4$ Department of Quantum Nanoscience, Kavli Institute of Nanoscience, Delft University of Technology, Delft 2628CJ, the Netherlands \\
    $^5$ Institut für Optik und Atomare Physik, Technische Universität Berlin, Straße des 17. Juni 135, 10623 Berlin, Germany \\
    $^6$ National Institute for Materials Science, Namiki 1-1, Tsukuba, 305-0044, Ibaraki, Japan \\[2ex]%
    \normalsize
  \vskip 1.5em
\end{titlepage}
\setcounter{figure}{0}
\section*{Tr-reflectivity measurements}
\begin{figure*}[h!]
\renewcommand\figurename{SI Figure}
    \centering
    \includegraphics[width=\linewidth, trim = 0mm 30mm 0mm 30mm,clip=true]{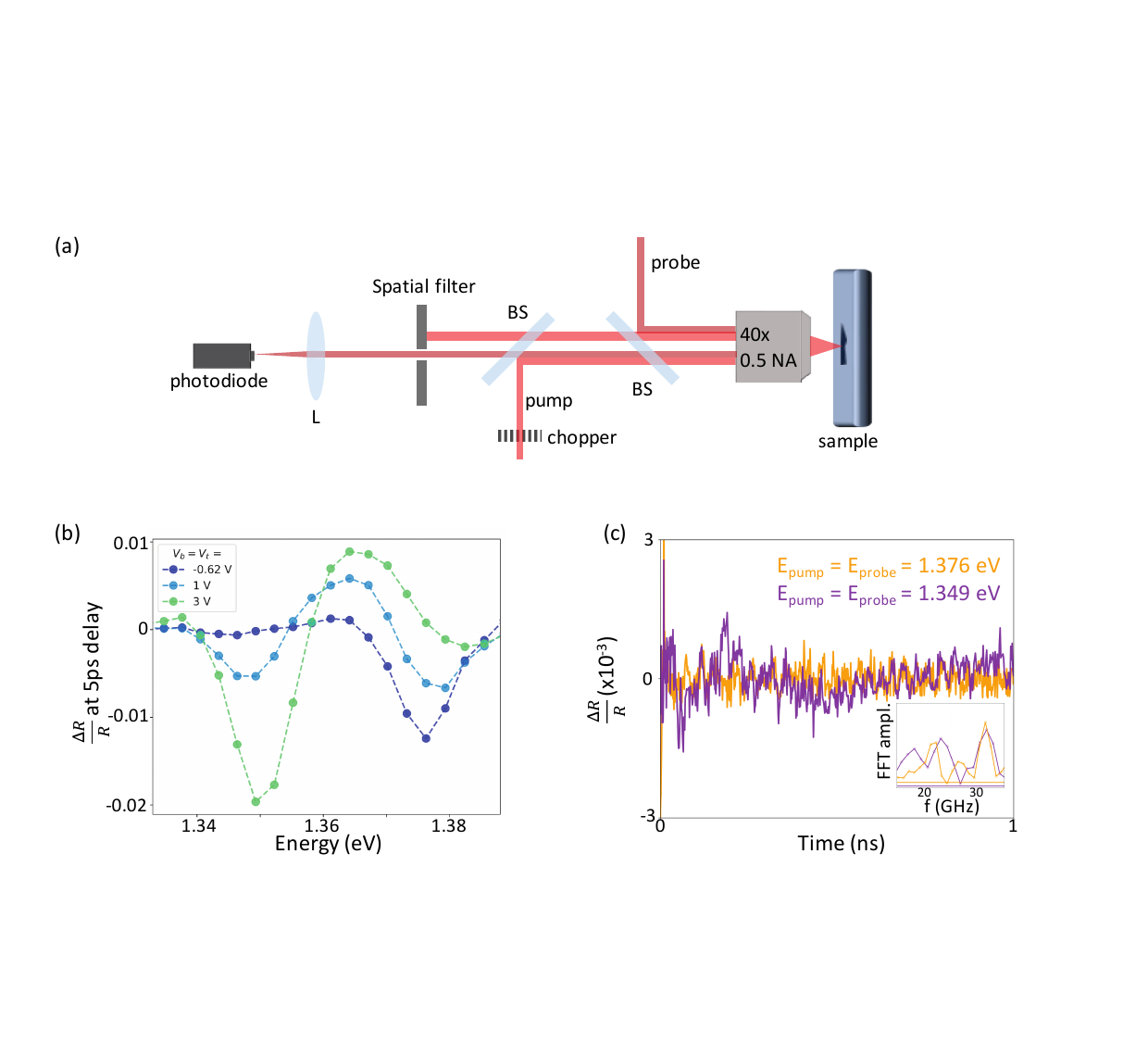}
    \caption{\textbf{Tr-reflecitivity measurement} (a) Detection scheme of single-color tr-reflectivity.
    (b) Tr-reflectivity spectra at small delay for different doping levels. The exciton resonance becomes weaker while the trion resonance strengthens for positive gate voltages.
    (c) Tr-reflectivity traces and extracted magnon frequencies (inset) measured at the exciton (orange) and trion (purple) resonances for $V_{\mathrm{b}} = V_{\mathrm{t}} = \SI{3}{\volt}$ after subtracting an exponential background.
     }
    \label{fig:SI1}
\end{figure*}

\begin{figure*}[h!]
\renewcommand\figurename{SI Figure}
    \centering
    \includegraphics[width=0.8\linewidth, trim = 0mm 20mm 0mm 50mm,clip=true]{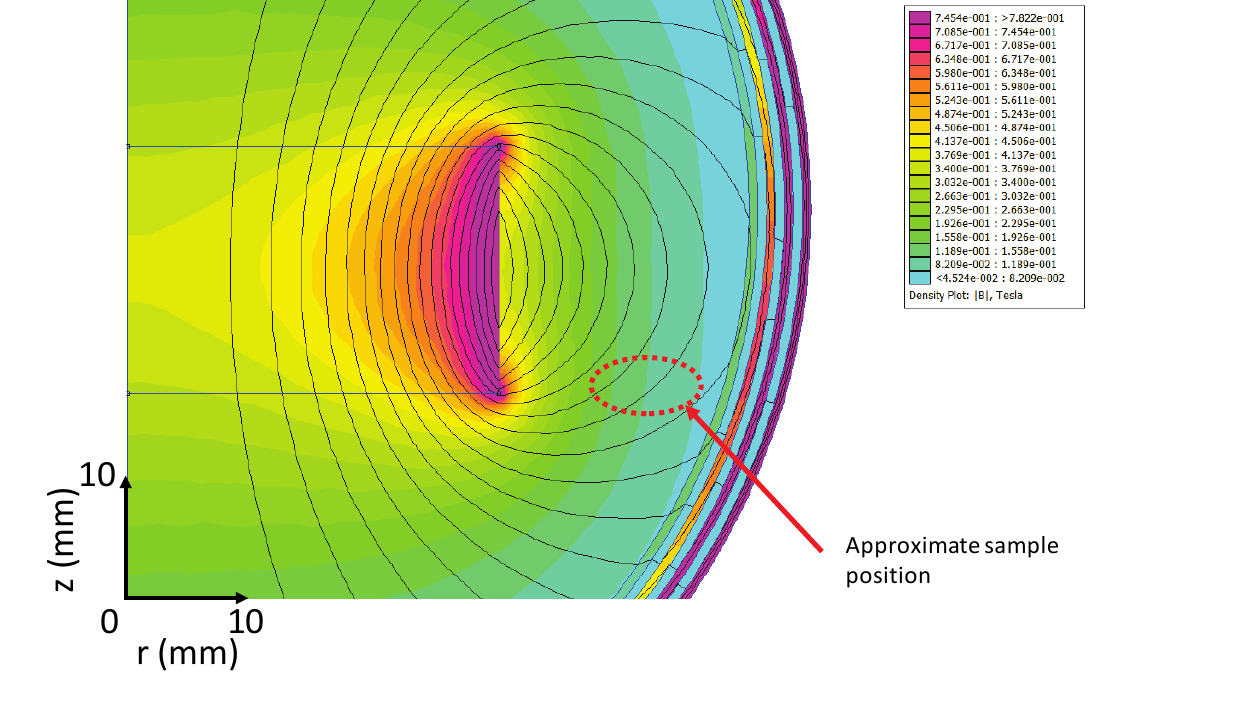}
    \caption{\textbf{Simulated field of permanent magnet used for experiments in the manuscript} Simulated magnetic field of the permanent magnet (Neodym 38, r = 3 mm) as a function of height $z$ and radial distance $r$.
     }
    \label{fig:SI10}
\end{figure*}

The single-color tr-reflectivity detection scheme is shown in \cref{fig:SI1}a. To achieve a high signal-to-noise ratio in our experiments, we tune our pump and probe energies just below the $X_{\mathrm{B}}$ exciton resonance around $\SI{1.375}{\eV}$ for low gate voltages (dark blue curve in \cref{fig:SI1}b). 
For higher gate voltages, it becomes neccessary to tune pump and probe to the trion resonance at $\SI{1.35}{\eV}$, as the exciton fades (green curve in \cref{fig:SI1}b). 
We see in \cref{fig:SI1}c, that for intermediate gate voltages the signal at both resonances (orange: exciton, purple: trion) becomes weak, however, they allow us to observe oscillations at both resonances and we find the same frequency for $f_{\mathrm{OP}}$ ($f_{\mathrm{IP}}$ ambiguous at trion resonance). \\
Additionally, we mount a permanent magnet in proximity to the sample to increase signal strength. To estimate the strength of the external magnetic field, we use the open source software FEMM4.2 \cite{femm} (\cref{fig:SI10}). At the sample position, the field is around $\SI{100}{\milli\tesla}$. It should be noted that the exact position of the sample can differ slightly between the measurements of the three devices, resulting in differences of the external magnetic field. Also, the external field has in- and out-of-plane components, as evidenced by the observation of both $f_{\mathrm{IP}}$ and $f_{\mathrm{OP}}$ modes. We neglect this fact in the macrospin model.
\newpage
\section*{Temperature- and fluence-dependent magnon frequency shifts}
\begin{figure*}[h]
\renewcommand\figurename{SI Figure}
    \centering
    \includegraphics[width=\linewidth, trim = 0mm 60mm 0mm 60mm,clip=true]{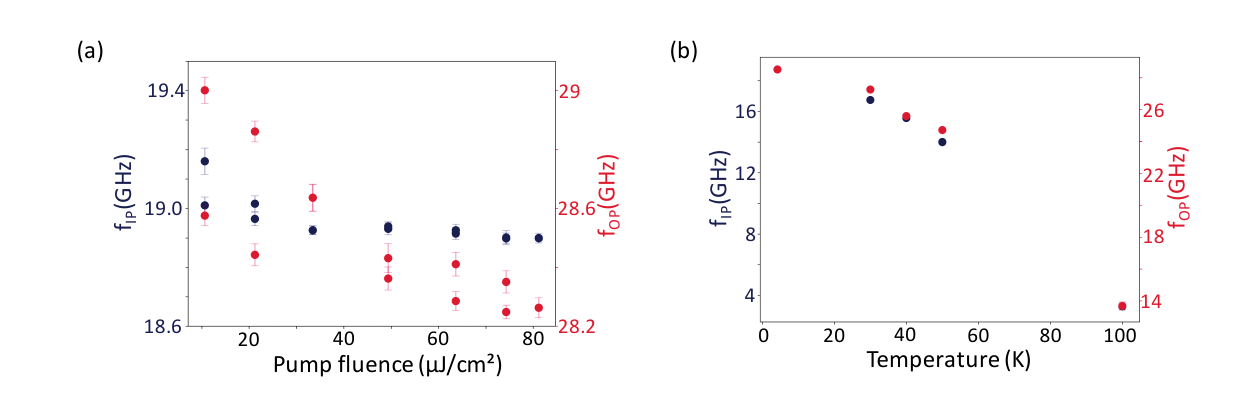}
    \caption{\textbf{Fluence- and temperature dependence of magnon modes} Shift of $f_{\mathrm{IP}}$ and $f_{\mathrm{OP}}$ as a function of (a) pump laser fluence and (b) sample temperature in the trilayer sample.
     }
    \label{fig:SI2}
\end{figure*}

To exclude laser- or gate-related heating effects as the origin of magnon frequency changes, fluence- and temperature dependent measurements were conducted on the trilayer sample. We see in \cref{fig:SI2}a that a higher fluence leads to a downshift in magnon frequencies of $\Delta f_{\mathrm{IP}} \approx - (0.1 - \SI{0.2}{\giga\hertz})$ and $\Delta f_{\mathrm{OP}} \approx - (0.4 - \SI{0.8}{\giga\hertz})$. This consistent with absorption-induced heating of the sample, as the magnon frequencies also shift down with temperature (\cref{fig:SI2}b). As shown in the main text, gating leads to an upshift of both magnon modes, so that we exclude heating as the underlying mechanism.
\newpage
\section*{Gate-dependent PL and electrostatic modeling}
\begin{figure*}[ht]
\renewcommand\figurename{SI Figure}
    \centering
    \includegraphics[width=\linewidth, trim = 0mm 0mm 0mm 0mm,clip=true]{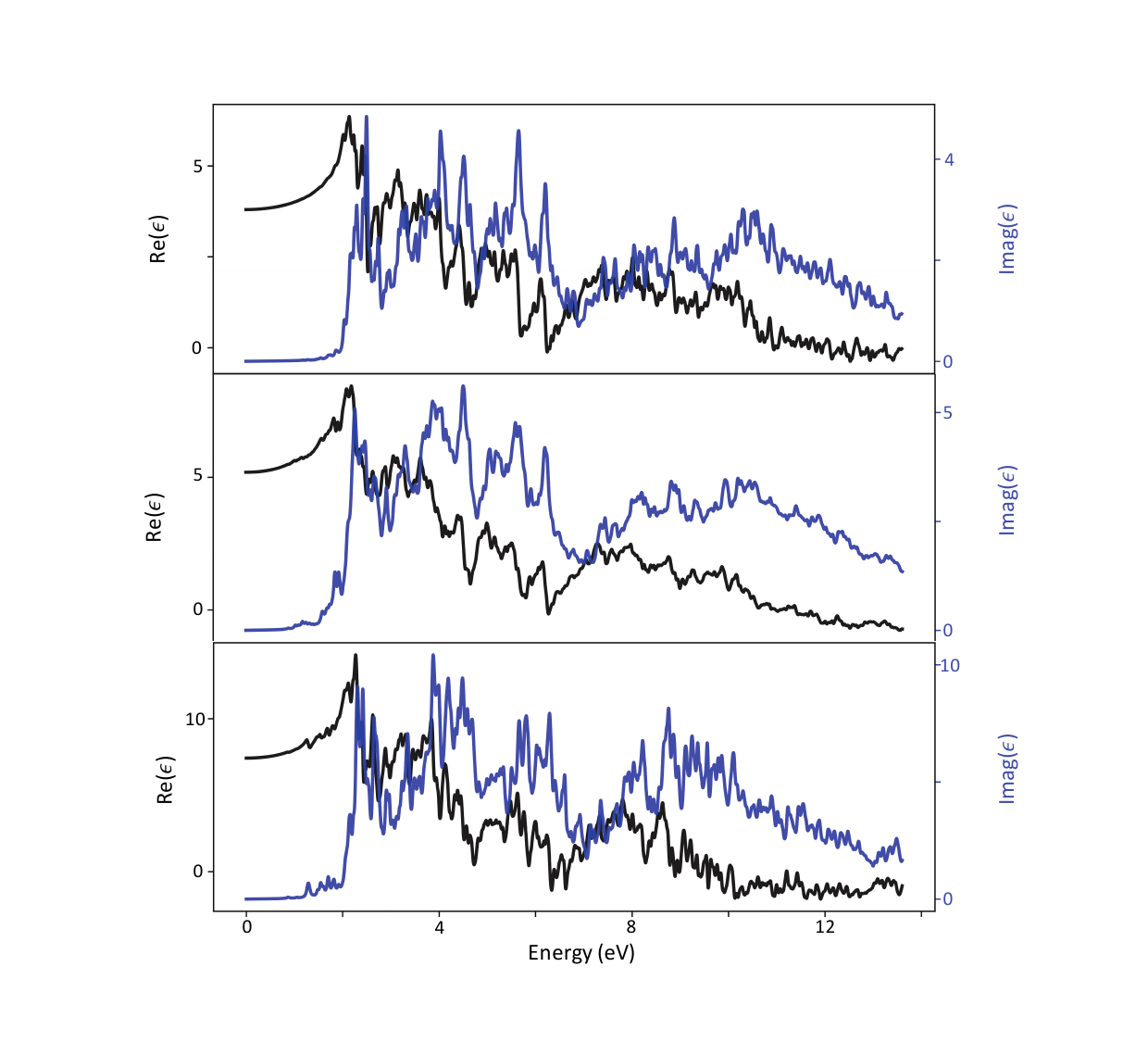}
    \caption{\textbf{Dielectric function of CrSBr} Calculated real and imaginary parts of the out-of-plane dielectric function of (a) monolayer, 
    (b) trilayer and (c) bulk CrSBr.
     }
    \label{fig:SI11}
\end{figure*}

\begin{figure*}[h!]
\renewcommand\figurename{SI Figure}
    \centering
    \includegraphics[width=\linewidth, trim = 0mm 0mm 0mm 0mm,clip=true]{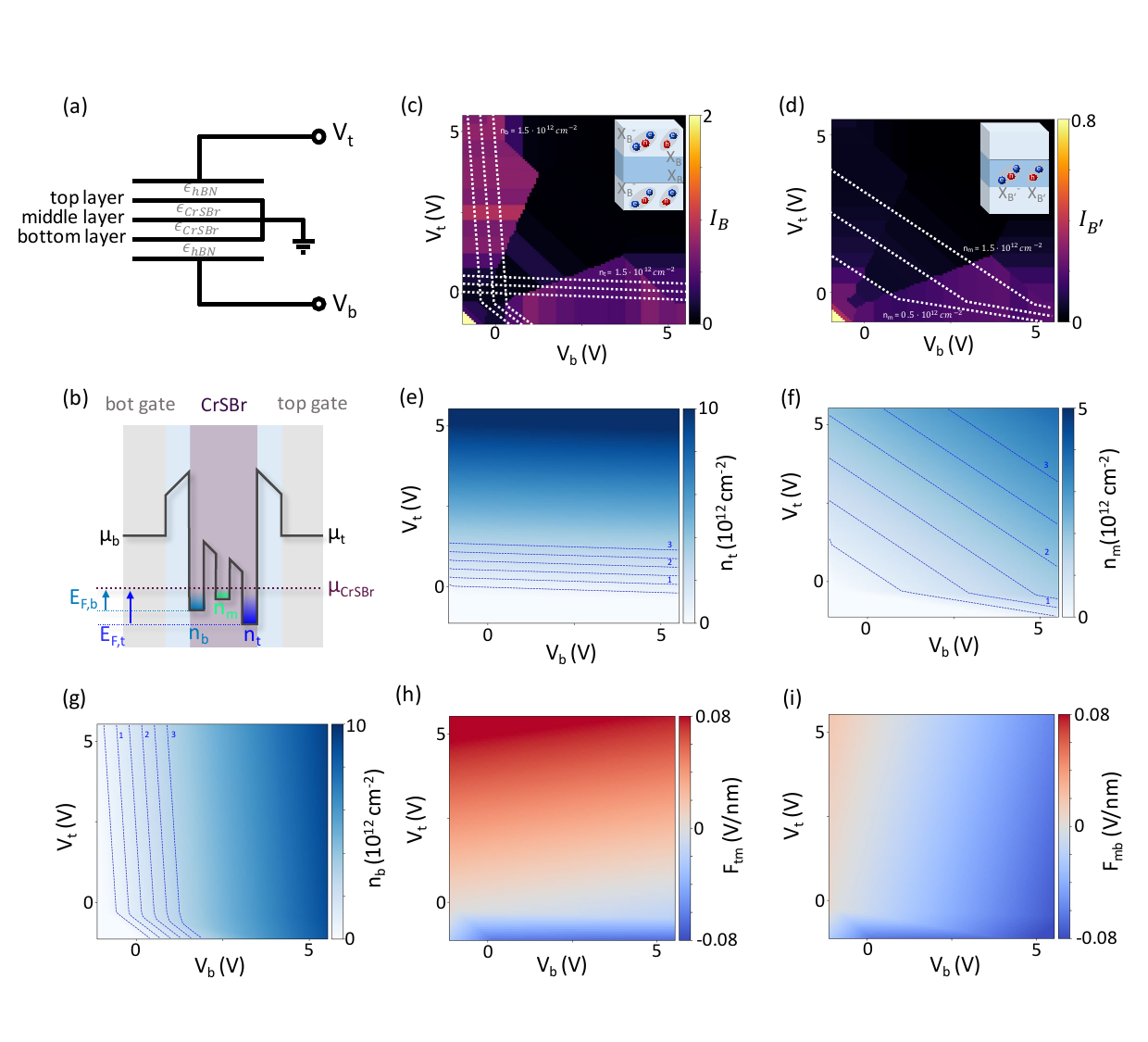}
    \caption{\textbf{Photoluminescence and electrostatic model of trilayer CrSBr} (a) Scheme of the capacitor model for a trilayer CrSBr with top and bottom gate and hBN dielectric.
    (b) Band alignment sketch when the sample is doped. The Fermi energies $E_{\mathrm{F},i}$ are defined as the difference between the conduction band and the chemical potential $\mu_{\mathrm{CrSBr}}$. The top and bottom gate voltages control $\mu_{\mathrm{t}}$ and $\mu_{\mathrm{b}}$. 
    (c) Exciton-to-trion ratio $I_{\mathrm{B}}$ for excitons and trions localized to top and bottom layer (see inset). The white dashed lines are guides to the eye corresponding to the contour lines for $n_{\mathrm{t}}$ and $n_{\mathrm{m}}$ being 0.5, 1 and  $\SI{1.5e12}{\per\centi\metre\squared}$ shown in (e), (g).
    (d) Exciton-to-trion ratio $I_{\mathrm{B'}}$ for excitons and trions localized in the middle layer (see inset). The white dashed lines are guides to the eye corresponding to the contour lines for $n_{\mathrm{m}} =$ 0.5, 1 and  $\SI{1.5e12}{\per\centi\metre\squared}$ shown in (f).
    (e-g) Electron density in the top (e), middle (f) and bottom (g) layer. The contour lines show densities of 0.5, 1, 1.5, 2, 2.5 and $\SI{3e12}{\per\centi\metre\squared}$.
    (h-i) Electric field between top and middle (h) and middle and bottom (i) layers.
     }
    \label{fig:SI3}
\end{figure*}

To quantify the layer-resolved electron densities and electric fields, we follow the capacitor model suggested in \cite{Pisoni2019,Bolotin2023}. 
In this model, the CrSBr layers correspond to capacitor plates separated by a dielectric with $\epsilon_{\mathrm{CrSBr}}$ and spaced by interlayer distance $d_{\mathrm{CrSBr}} = \SI{0.8}{\nano\metre}$. Using density functional theory (DFT), we calculate the out-of-plane $\epsilon_{\mathrm{CrSBr}} (E)$ for mono-, trilayer and bulk, shown in \cref{fig:SI11}. As DC dielectric constants ($E = 0$), we find $\epsilon_{\mathrm{CrSBr}} = 3.8$ for a monolayer, $\epsilon_{\mathrm{CrSBr}} = 5.19$ for the trilayer and $\epsilon_{\mathrm{CrSBr}} = 7.42$ for the bulk. The graphite gates are modelled as capacitor plates with a dielectric of $\epsilon_{\mathrm{hBN}} = 3.76$ \cite{Laturia2018} and a thickness of the hBN flakes $d_{\mathrm{hBN}}$.
The equivalent circuit for the capacitor model for the trilayer is shown in \cref{fig:SI3}a, and the resulting energy diagram in \cref{fig:SI3}b.
In the trilayer sample, the hBN thicknesses on both sides are $d_{\mathrm{hBN}} = \SI{10}{\nano\metre}$. 
\\
The system of equations to extract the Fermi level with respect to the bottom of the conduction band $E_{F,i}$ (in eV, see \cref{fig:SI3}b) and the carrier density $n_i$ (where $i$ is top, middle and bottom) in the trilayer is as follows:
\begin{align}
    0 &= (V_{\mathrm{t}} + V_{\mathrm{t,}0}) - E_{\mathrm{F,t}} - \frac{en_{\mathrm{t}}}{C_{\mathrm{t}}} - \frac{C_{\mathrm{CrSBr}}}{C_{\mathrm{t}}}(E_{\mathrm{F,t}}-E_{\mathrm{F,m}})\\
    0 &= (V_{\mathrm{b}} + V_{\mathrm{b,}0}) - E_{\mathrm{F,b}} - \frac{en_{\mathrm{b}}}{C_{\mathrm{t}}} - \frac{C_{\mathrm{CrSBr}}}{C_{\mathrm{b}}}(E_{\mathrm{F,b}}-E_{\mathrm{F,m}})\\
    0 &= -en_{\mathrm{m}} - \frac{C_{\mathrm{CrSBr}}}{e}(E_{\mathrm{F,m}}-E_{\mathrm{F,t}}) - \frac{C_{\mathrm{CrSBr}}}{e}(E_{\mathrm{F,m}}-E_{\mathrm{F,b}}).
    \label{eq:SIcapacitor3L}
\end{align}
Here $C_{\mathrm{CrSBr}} = \frac{\epsilon_0\epsilon_{\mathrm{CrSBr}}}{d_{\mathrm{CrSBr}}}$ and $C_{\mathrm{t/b}} = \frac{\epsilon_0\epsilon_{\mathrm{hBN}}}{d_{hBN,t/b}}$.
To calculate the electron densities, we assume a two-dimensional density of states (DOS) similar to that in TMDs \cite{Bolotin2023}. We note that behavior suggesting a one-dimensional DOS has been reported in CrSBr \cite{Klein2023}, which would lead to a stronger increase in $n$ for energies just above the conduction band due to the emergence of van't Hove singularities. However, the overall dependence of $n$ and $F$ on the gate voltages, and therefore the main results of the manuscript, will not change with a different DOS. 
\\
In the PL data of the main manuscript we observe two key features which are of importance for the electrostatic modeling: Our CrSBr crystals are intrinsically n-doped (signified by the dominance of trions over excitons at zero gate voltage) and our samples have a built-in electric field (as shown by the field dependence of the middle layer exciton-to-trion ratio). To model the intrinsic n-doping, we introduce positive offset voltages $V_{\mathrm{t,}0}$ and $V_{\mathrm{b,}0}$ in \cref{eq:SIcapacitor3L} -- corresponding to a downshift of the conduction bands of all layers with respect to the vacuum level. To model the built-in electric field, we use $V_{\mathrm{t,}0} \neq V_{\mathrm{b,}0}$ to introduce an offset between the conduction bands across layers. \\
To find reasonable values for $V_{\mathrm{t,}0}$ and $V_{\mathrm{b,}0}$, we use the exciton-to-trion ratios $I_{\mathrm{B}} = \frac{I_{X_{\mathrm{B}}}}{I_{X_{\mathrm{B}}^-}}$ and $I_{\mathrm{B'}} = \frac{I_{X_{\mathrm{B}}'}}{I_{X_{\mathrm{B}}'^-}}$ as indicators for the carrier densities of the outer layers, $n_{\mathrm{t}}+n_{\mathrm{b}}$, and the middle layer, $n_{\mathrm{m}}$. As it is difficult to directly correlate $I_{\mathrm{B}}$ and $I_{\mathrm{B'}}$ to the electron densities by fitting (especially as we cannot 'resolve' $n_{\mathrm{t}}$ and $n_{\mathrm{b}}$ separately in the PL), we manually vary the offset voltages in the model to match the onset of doping to drops in the exciton-to-trion ratios. \\
From the symmetric behavior of $I_{\mathrm{B}}$ in \cref{fig:SI3}c we deduce the following:
\begin{itemize}
    \item We see the highest exciton-to-trion ratio $I_{\mathrm{B}}$ when both $V_{\mathrm{b}}$ and $V_{\mathrm{t}}$ $< \SI{-0.8}{\volt}$. This means that increasing either gate voltage above this value should result in at least one of the layers being electron doped.
    \item When varying only one of the gate voltages while fixing the other to a sufficiently negative value, either the top or bottom layer stays almost undoped, as $I_{\mathrm{B}}$ stays nearly constant after an initial drop.
    \item When both $V_{\mathrm{b}}$ and $V_{\mathrm{t}}$ $> \SI{0}{\volt}$, both top and bottom layers should be electron-doped, as $I_{\mathrm{B}} \approx 0$ for those ranges of the gating diagram.
\end{itemize}
This gives us an upper limit for $V_{\mathrm{t,}0}$ and $V_{\mathrm{b,}0}$ of $\sim \SI{2.5}{\volt}$, as increasing the offset voltages beyond this limit would mean that the sample is doped in the top or bottom layer for all gate voltages we reach in the experiment.\\
In the gate-dependence of $I_{\mathrm{B'}}$ in \cref{fig:SI3}d, we see that
\begin{itemize}
    \item $I_{\mathrm{B'}}$ drops faster when applying a top gate voltage. From this, we infer that $V_{\mathrm{t,}0} < V_{\mathrm{b,}0}$, because a smaller $V_{\mathrm{t}}$ is then sufficient to dope the middle layer.
\end{itemize} 
We find that $V_{\mathrm{t,}0} = \SI{1}{\volt}$ and $V_{\mathrm{b,}0} = \SI{1.7}{\volt}$ fulfill all the requirements mentioned above and adequately reproduce the behavior seen in PL. The resulting layer-resolved doping maps are shown in \cref{fig:SI3}e-g, with contour lines for selected doping levels ($0.5- \SI{3e12}{\per\centi\metre\squared}$). \\
We show the same contour lines as guides to the eye in the experimental $I_{\mathrm{B}}$ and $I_{\mathrm{B'}}$ maps in \cref{fig:SI3}c, d. In \cref{fig:SI3}c, we see that the contour lines show the same symmetry with gates as $I_{\mathrm{B}}$, and we see a drop in $I_{\mathrm{B}}$ for a similar $n_{\mathrm{t}}$ or $n_{\mathrm{b}}$ (around $\sim \SI{1.5e12}{\per\centi\metre\squared}$). In \cref{fig:SI3}d, the asymmetry of the contour line for $n_{\mathrm{m}} = \SI{0.5e12}{\per\centi\metre\squared}$ shows that we need a higher $V_{\mathrm{b}}$ to dope the middle layer when $V_{\mathrm{t}} < 0$ than in the opposite case. The modeled $n_{\mathrm{m}}$ also shows a quick increase when increasing $V_{\mathrm{t}}$ for large enough $V_{\mathrm{b}}$, concurrent with a drop in $I_{\mathrm{B'}}$ in the same case. While the onset of the $n_{\mathrm{m}}$ happens for slightly lower gate voltages than the drop in $I_{\mathrm{B'}}$, the offset voltages of $V_{\mathrm{t,}0} = \SI{1}{\volt}$ and $V_{\mathrm{b,}0} = \SI{1.7}{\volt}$ give the best overall correlation between modeling and PL data. The gate-dependent electric fields between the layers are shown in \cref{fig:SI3}h, i. \\
It should be noted that changing the offset voltages slightly does not change the overall predicted behavior of the frequencies in the macrospin model (i.e. the gate sensitivity or frequency range). It can, however, slightly change the onset of frequency shifts as well as the fitted $\nu_{\mathrm{a}}$, $\nu_{\mathrm{E}}$ and $\eta_{\gamma}$.
\section*{Gate-dependent PL and electrostatic model of 5-layer device}
\begin{figure*}[h!]
\renewcommand\figurename{SI Figure}
    \centering
    \includegraphics[width=\linewidth, trim = 0mm 25mm 0mm 55mm,clip=true]{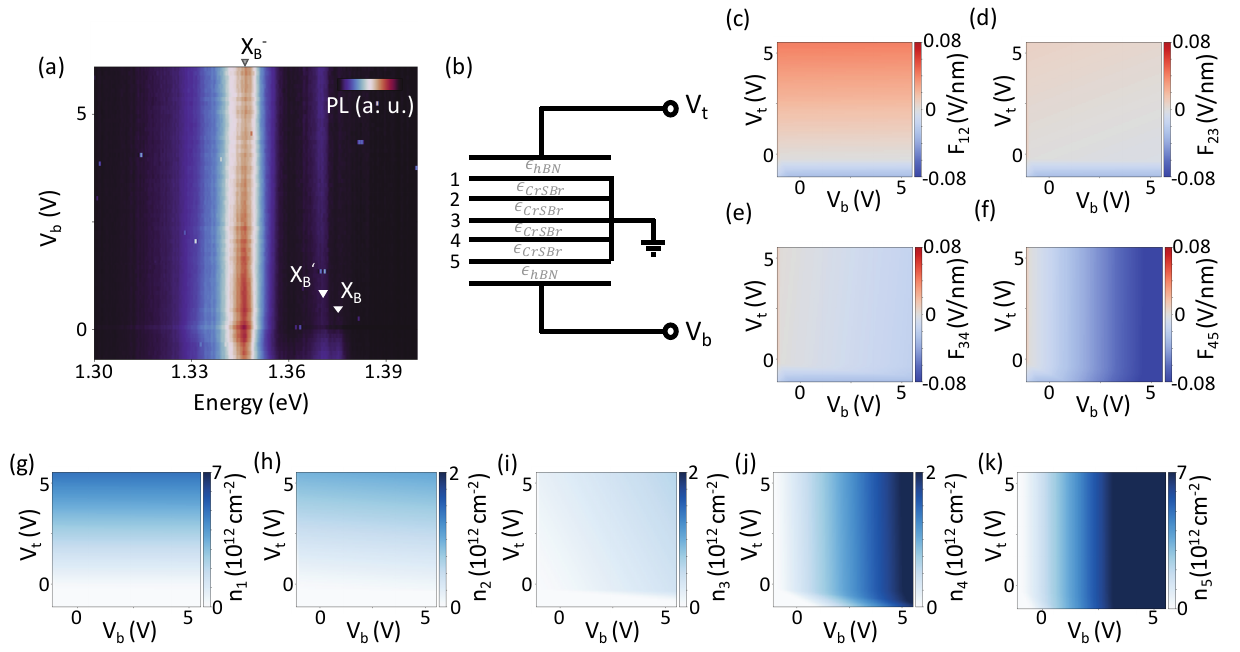}
    \caption{\textbf{Photoluminescence and electrostatic model of 5-layer CrSBr} (a) $V_{\mathrm{b}}$-dependent photoluminescence map of the 5-layer CrSBr device. 
    (b) Scheme of the capacitor model for a 5-layer CrSBr with top and bottom gate and hBN dielectric.
    (c-f) Electric field between the respective layers (numbering see (b)).
    (g-k) Electron density in the respective layers (numbering see (b)).
     }
    \label{fig:SI4}
\end{figure*}     
The gate-dependent photoluminescence of the 5-layer device in \cref{fig:SI4}a again shows two split excitonic peaks at $X_{\mathrm{B}} \approx \SI{1.376}{\eV}$ and $X_{\mathrm{B}}' \approx \SI{1.370}{\eV}$, which we hypothesize to reside in the outer and inner layers, respectively. The intensities of the excitonic peaks for the undoped sample ($V_{\mathrm{b}} < \SI{0}{\volt}$) are now comparable, which we attribute to the increased number of layers hosting $X_{\mathrm{B}}'$ excitons. In this device we could only apply a bottom gate voltage. For $V_{\mathrm{b}} > \SI{0}{\volt}$, the intensity of the $X_{\mathrm{B}}$ exciton drops, and the trion resonance around $\SI{1.345}{\eV}$ brightens. Meanwhile, the $X_{\mathrm{B}}'$ exciton remains visible for all measured gate voltages. \\
When modeling the 5-layer device as shown in \cref{fig:SI4}b, we expand the system of equations to
\begin{align}
        0 &= (V_{\mathrm{t}} + V_{\mathrm{t,}0}) - E_{\mathrm{F,1}} - \frac{en_{1}}{C_{\mathrm{t}}} - \frac{C_{\mathrm{CrSBr}}}{C_{\mathrm{t}}}(E_{\mathrm{F,1}}-E_{\mathrm{F,2}})\\
        0 &= (V_{\mathrm{b}} + V_{\mathrm{b,}0}) - E_{\mathrm{F,5}} - \frac{en_{5}}{C_{\mathrm{b}}} - \frac{C_{\mathrm{CrSBr}}}{C_{\mathrm{b}}}(E_{\mathrm{F,5}}-E_{\mathrm{F,4}})\\
        0 &= -en_2 - \frac{C_{\mathrm{CrSBr}}}{e}(E_{\mathrm{F,2}}-E_{\mathrm{F,1}}) - \frac{C_{\mathrm{CrSBr}}}{e}(E_{\mathrm{F,2}}-E_{\mathrm{F,3}})\\
        0 &= -en_3 - \frac{C_{\mathrm{CrSBr}}}{e}(E_{\mathrm{F,3}}-E_{\mathrm{F,2}}) - \frac{C_{\mathrm{CrSBr}}}{e}(E_{\mathrm{F,3}}-E_{\mathrm{F,4}})\\
        0 &= -en_4 - \frac{C_{\mathrm{CrSBr}}}{e}(E_{\mathrm{F,4}}-E_{\mathrm{F,3}}) - \frac{C_{\mathrm{CrSBr}}}{e}(E_{\mathrm{F,4}}-E_{\mathrm{F,5}})
\end{align}
The hBN thicknesses in the 5-layer device are $d_{hBN,top} = \SI{5}{\nano\metre}$ and $d_{hBN,bot} = \SI{10.5}{\nano\metre}$. We use the trilayer $\epsilon_{\mathrm{CrSBr}} = 5.19$ calculated with DFT (\cref{fig:SI11}b). We assume the same offset voltages $V_{\mathrm{t,}0}$ and $V_{\mathrm{b,}0}$ as for the trilayer device. This leads to $n_4$ and $n_5$ increasing around $V_{\mathrm{b}} \approx -\SI{0.5}{\volt}$, in good agreement with the drop in $I_{\mathrm{B'}}$ shortly after.
We also see that the charge carrier densities in the center layer remain below $\approx \SI{1e12}{\per\centi\metre\squared}$ for all applied gate voltages (\cref{fig:SI4}i), which agrees with the visibility of $X_{\mathrm{B}}'$ for all gate voltages.
All the calculated fields and electron densities for the 5-layer device are shown in \cref{fig:SI4}c-k.
\section*{Gate-dependent PL and electrostatic model of 8-layer device}
\begin{figure*}[h!]
\renewcommand\figurename{SI Figure}
    \centering
    \includegraphics[width=\linewidth, trim = 0mm 0mm 0mm 0mm,clip=true]{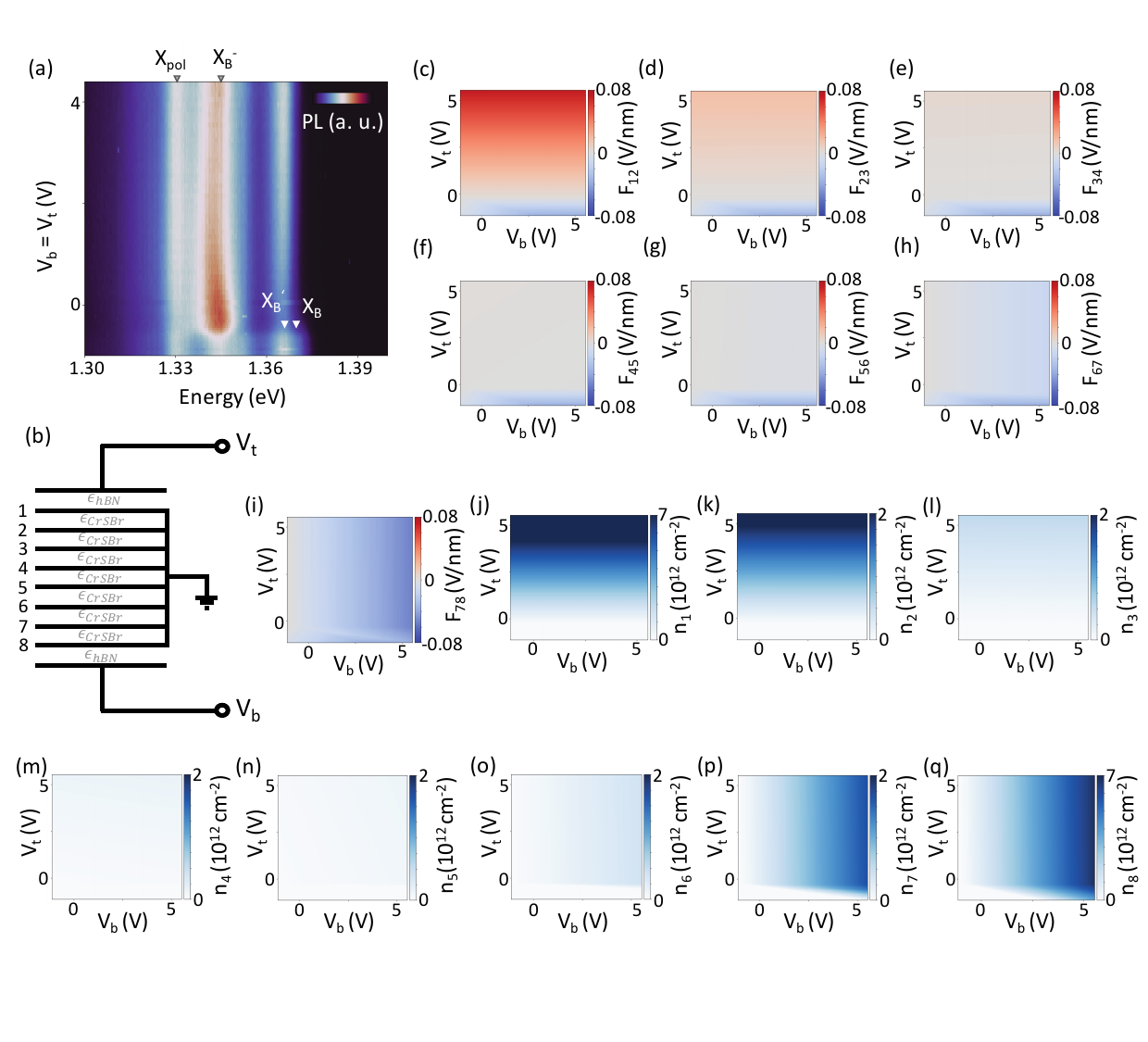}
    \caption{\textbf{Photoluminescence and electrostatic model of 8-layer CrSBr} (a) Gate-dependent photoluminescence map of the 8-layer CrSBr device. 
    (b) Scheme of the capacitor model for an 8-layer CrSBr with top and bottom gate and hBN dielectric.
    (c-h) Electric field between the respective layers (numbering see (b)).
    (i-q) Electron density in the respective layers (numbering see (b)).
     }
    \label{fig:SI5}
\end{figure*}     
In the doping-dependent photoluminescence of the 8-layer device in \cref{fig:SI5}a, we once again see two split excitonic peaks at $X_{\mathrm{B}} \approx \SI{1.372}{\eV}$ and $X_{\mathrm{B}}' \approx \SI{1.366}{\eV}$. In contrast to the tri- and 5-layer device, the $X_{\mathrm{B}}'$ peak is more intense than $X_{\mathrm{B}}$, the latter only visible as a high-energy shoulder. This is consistent with $X_{\mathrm{B}}'$ emission stemming from excitons localized to the inner layers, as the number of such inner layers has increased from the 3- to 8-layer device. \\
We see $X_{\mathrm{B}}$ disappear for $V_{\mathrm{b}} = V_{\mathrm{t}} > \SI{-0.3}{\volt}$, simultaneously with a brightening of the trion resonance around $\SI{1.345}{\eV}$. The $X_{\mathrm{B}}'$ exciton remains visible for all measured gate voltages, similarly to the 5-layer device.
Additional peaks $X_{pol}$ are visible in the 8-layer device PL at lower energies, possibly stemming from polaritons due to the increased sample thickness \cite{Dirnberger2022}. \\
We expand the electrostatic model to 8 layers (\cref{fig:SI5}b) in the same manner as for 5 layers.
We measure the hBN thicknesses to be $d_{hBN,top} = \SI{10}{\nano\metre}$ and $d_{hBN,bot} = \SI{6}{\nano\metre}$ with AFM. Due to the increased thickness of the device, we use the bulk dielectric constant  $\epsilon_{\mathrm{CrSBr}} = 7.42$ calculated with DFT (\cref{fig:SI11}c). We again assume the same offset voltages $V_{\mathrm{t,}0}$ and $V_{\mathrm{b,}0}$ used for the other devices. The resulting increase of electron density in the bottom layer $n_8$ around $V_{\mathrm{b}} = V_{\mathrm{t}} \approx -\SI{0.4}{\volt}$ (\cref{fig:SI5}q), followed by the top layer $n_1$ around $V_{\mathrm{b}} = V_{\mathrm{t}} \approx -\SI{0.2}{\volt}$ correlates well with the drop in the exciton intensity we observe at $V_{\mathrm{b}} = V_{\mathrm{t}} > \SI{-0.3}{\volt}$. The middle layers of the 8-layer device remain almost undoped (\cref{fig:SI5}m, n).
All the calculated fields and electron densities for the 8-layer device are shown in \cref{fig:SI5}c-q.
\section*{Bulk macrospin model}
\begin{figure*}[h]
\renewcommand\figurename{SI Figure}
    \centering
    \includegraphics[width=\linewidth, trim = 20mm 50mm 20mm 50mm,clip=true]{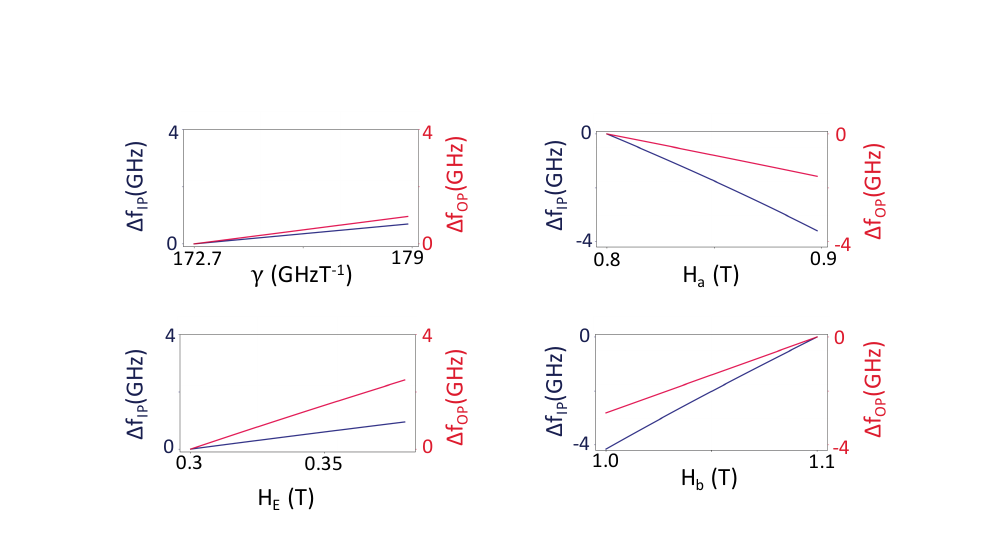}
    \caption{\textbf{Analytical macrospin model} Influence of changing (a) $\gamma$, (b) $H_{\mathrm{E}}$, (c) $H_{\mathrm{a}}$ and (d) $H_{\mathrm{b}}$ on the magnon modes.
     }
    \label{fig:SI12}
\end{figure*}

For bulk antiferromagnets, we assume two coupled macrospins which repeat itself in periodic boundary conditions. We can write the LL equation
\begin{equation}
    \frac{d\vec{m}_{1,2}}{dt} = -\gamma \vec{m}_{1,2}\times\vec{H}_{\mathrm{eff}_{1,2}}
    \label{eq:SILLG}
\end{equation}
where $\vec{m}_{1,2}$ are the macrospins in neighboring layers. The effective field $\vec{H}_{\mathrm{eff}_{1,2}}$ can be obtained from the energy $E$ of the system:
\begin{align}
    E &= H_0\hat{c}\cdot(\vec{m}_1+\vec{m}_2) - H_{\mathrm{E}}\vec{m}_{1}\cdot\vec{m}_2 + \frac{1}{2}H_{\mathrm{a}} ((m_{1}^a)^2 + (m_{2}^a)^2) + \frac{1}{2}H_{\mathrm{b}} ((m_{1}^b)^2 + (m_{2}^b)^2) \\
    \vec{H}_{\mathrm{eff}_{1,2}} &= -\nabla_{\vec{m}_{1,2}}E = H_0\hat{c} - H_{\mathrm{E}}\vec{m}_{2,1} + H_{\mathrm{a}} m_{1,2}^a \hat{a} + H_{\mathrm{b}} m_{1,2}^b \hat{b} .
    \label{eq:SIefffield}
\end{align}
Here, $\hat{a}, \hat{b}, \hat{c}$ point along the corresponding crystallographic axes of CrSBr. $H_{\mathrm{b}}$ describes the easy axis magnetic anisotropy and $H_{\mathrm{a}}$ the intermediate one. $H_{\mathrm{E}}$ is the interlayer exchange interaction, which aligns the neighboring macrospins antiparallely. The external magnetic field $\vec{H}_{0} = H_0 \hat{c}$ is assumed out-of-plane and determines the initial tilt $\theta$ of the macrospins away from the easy $\hat{b}$-axis. \\
As $H_{0}$ does not saturate the spins along the $\hat{c}$-axis, the equilibrium macrospins can be written as $\vec{m}_{{1,2}_0} = \left( 0, \sin \theta, \pm \cos \theta \right)$. By inserting this expression into \cref{eq:SIefffield}, we find $\theta$ by minimizing the equilibrium energy, $\frac{\mathrm{d}E}{\mathrm{d}\theta} = 0$, leading to the condition
\begin{equation}
    H_0 = \sin \theta (H_{\mathrm{b}}+2H_{\mathrm{E}}). 
    \label{eq:SIangle_eq}   
\end{equation}
Next, we express the macrospins as $\vec{m}_{1,2} = \vec{m}_{{1,2}_0} + \delta\vec{m}_{1,2}$, where $\delta\vec{m}_{1,2} << \vec{m}_{{1,2}_0}$ are small deviations from equilibrium after excitation. 
By linearizing the LL equation, we arrive at 
\begin{equation}
    \frac{\mathrm{d}}{\mathrm{d}t}\delta\vec{m}_{1,2} = -\gamma \left(\delta\vec{m}_{1,2}\times\vec{H}_{\mathrm{eff}_{1,2}}^{0} + \vec{m}_{{1,2}_0}\times\delta \vec{H}_{eff_{1,2}} \right)
    \label{eq:SIlinearizedLLG}
\end{equation}
where $\vec{H}_{\mathrm{eff}_{1,2}} = \vec{H}_{\mathrm{eff}_{1,2}}^{0} + \delta\vec{H}_{\mathrm{eff}_{1,2}}$. 
\\
From the condition $\vec{m}_{{1,2}_0}\cdot\delta\vec{m}_{1,2} = 0$, we obtain $\delta\vec{m}_{1,2} = \left( \delta m_{1,2}^{\mathrm{a}}, \mp \delta m_{1,2}^{||} \cos \theta, \delta m_{1,2}^{||} \sin \theta\right)$. 
Inserting this into \cref{eq:SIlinearizedLLG} 
and using the ansatz $\delta \vec{m}_{1,2} (t) = \delta \vec{m}_{1,2} (t=0) e^{-i\omega t}$ results in 
\begin{equation}
    \frac{i\omega}{\gamma} \begin{pmatrix}
        \delta m_1^a \\ \delta m_1^{||} \\ \delta m_2^a \\ \delta m_2^{||}
    \end{pmatrix} = \begin{pmatrix}
        0 & -H_{\mathrm{E}} -H_{\mathrm{b}}\cos^2 \theta & 0 & H_{\mathrm{E}}\cos (2\theta) \\
        H_{\mathrm{E}} + H_{\mathrm{b}} - H_{\mathrm{a}} & 0 & H_{\mathrm{E}} & 0 \\
        0 & H_{\mathrm{E}} \cos (2\theta) & 0 & -H_{\mathrm{E}}-H_{\mathrm{b}}\cos^2\theta \\
        H_{\mathrm{E}} & 0 & H_{\mathrm{E}} + H_{\mathrm{b}} - H_{\mathrm{a}} & 0 
    \end{pmatrix}\begin{pmatrix}
        \delta m_1^a \\ \delta m_1^{||} \\ \delta m_2^a \\ \delta m_2^{||}
    \end{pmatrix}.
    \label{eq:SImatrix_analyticmodel}
\end{equation}
Solving this equation gives an analytical solution for the eigenvalues:
\small
\begin{equation}
    \begin{split}
    f_{\mathrm{IP}} &= \pm \frac{\gamma}{2\pi}\sqrt{(H_{\mathrm{b}}-H_{\mathrm{a}})(H_{\mathrm{b}}+2H_{\mathrm{E}})\left(1-\frac{H_0^2}{(2H_{\mathrm{E}}+H_{\mathrm{b}}-H_{\mathrm{a}})^2}\right)} \\
    f_{\mathrm{OP}} &= \pm \frac{\gamma}{2\pi}\sqrt{(H_{\mathrm{b}}-H_{\mathrm{a}}+2H_{\mathrm{E}})(H_{\mathrm{b}}+2H_{\mathrm{E}})\left(2H_{\mathrm{E}}\frac{H_0^2}{(2H_{\mathrm{E}}+H_{\mathrm{b}}-H_{\mathrm{a}})^2}-H_{\mathrm{b}}\left(\frac{H_0^2}{(2H_{\mathrm{E}}+H_{\mathrm{b}}-H_{\mathrm{a}})^2}-1\right)\right)}.
    \end{split}
    \label{eq:SIanalyticfreq}
\end{equation} 
\normalsize
As the ansatz assumes that every second macrospin is exactly the same, these two solutions correspond to the case of zero momentum magnons (i.e. $f_{\mathrm{IP}}$ and $f_{\mathrm{OP}}$ in bulk experiments). \\
For magnons with $k_{\mathrm{z}} \neq 0$, however, there is a phase difference between every second macrospin. Including this into the solutions of \cref{eq:SIlinearizedLLG} gives an analytical solution for the dispersion along $k_z$. The phase difference changes the interlayer exchange term in the effective field of each macrospin $\vec{m}_i$ (where $i$ is the layer number), since the macrospins above and below are no longer exactly the same. The effective field in layer $i$ becomes
\begin{align}
    \vec{H}_{\mathrm{eff}_{i}} &= H_0\hat{c} - \frac{1}{2}H_{\mathrm{E}}(\vec{m}_{i-1} + \vec{m}_{i+1}) + H_{\mathrm{a}} m_{i}^a \hat{a} + H_{\mathrm{b}} m_{i}^b \hat{b}.
    \label{eq:SIeff_field_dispersion}
\end{align}
To solve the new LL equation, we assume the following: Before the excitation, the equilibrium macrospins in every second layer are the same, $\vec{m}_{{i+1}_0} = \vec{m}_{{i-1}_0}$. For the deviations after the excitations, we use a plane wave ansatz $\delta \vec{m}_j (t) = \delta \vec{m}_j (t=0) e^{-i\omega t}e^{ik_zjd_z}$, where $d_z$ is the interlayer spacing in z-direction and $j$ describes the layer number. \\
Inserting the new exchange term from \cref{eq:SIeff_field_dispersion} in \cref{eq:SIlinearizedLLG} gives rise to the terms
\begin{equation}
    \sim H_{\mathrm{E}}\begin{pmatrix}
        \delta m_i^{||}(\cos ^2 \theta - \sin ^2 \theta) \\ \delta m_i^{a}\cos\theta \\ \delta m_i^{\mathrm{a}} \sin\theta
    \end{pmatrix} - H_{\mathrm{E}}\begin{pmatrix}
        \delta m_{i-1}^{||}(\sin ^2 \theta - \cos ^2 \theta) \\ \delta m_{i-1}^{a}\cos\theta \\ -\delta m_{i-1}^a \sin\theta
    \end{pmatrix} (1+e^{-2ik_zd_z})
\end{equation}
which are the same as the ones found for $k_z = 0$, apart from the last term $\propto e^{-2ik_zd_z}$.
This changes the matrix equation from \cref{eq:SImatrix_analyticmodel} to:
\small
\begin{equation}
    \frac{i\omega}{\gamma} \begin{pmatrix}
        \delta m_1^a \\ \delta m_1^{||} \\ \delta m_2^a \\ \delta m_2^{||}
    \end{pmatrix} = 
    \begin{pmatrix}
        0 & -H_{\mathrm{E}} -H_{\mathrm{b}}\cos^2 \theta & 0 & H_{\mathrm{E}}\cos (2\theta) \left(1+e^{-2ik_z\cdot d_z}\right)\\
        H_{\mathrm{E}} + H_{\mathrm{b}} - H_{\mathrm{a}} & 0 & H_{\mathrm{E}}\left(1+e^{-2ik_z\cdot d_z}\right) & 0 \\
        0 & H_{\mathrm{E}} \cos (2\theta)\left(1+e^{-2ik_z\cdot d_z}\right) & 0 & -H_{\mathrm{E}}-H_{\mathrm{b}}\cos^2\theta \\
        H_{\mathrm{E}}\left(1+e^{-2ik_z\cdot d_z}\right) & 0 & H_{\mathrm{E}} + H_{\mathrm{b}} - H_{\mathrm{a}} & 0 
    \end{pmatrix}
    \begin{pmatrix}
        \delta m_1^a \\ \delta m_1^{||} \\ \delta m_2^a \\ \delta m_2^{||}
    \end{pmatrix}.
    \label{eq:SImatrix_analyticmodel_planewave}
\end{equation}
\normalsize
We solve this to extract the dispersion relation in Fig. 3a of the main text.\\
In \cref{fig:SI12} we show the dependance of the modes on gyromagnetic ratio, anisotropies and interlayer exchange. They mirror those shown in Fig. 3 of the main text for the numerical model. We also see that increasing $H_a$ and decreasing $H_b$ has a similar effect on the frequencies, and thus we fix $H_b$ for all fits to avoid overfitting. 
\section*{Layer-resolved macrospin model}
For the layer-resolved macrospin model, we again introduce a layer-dependent effective field into the LL equation. Now, however, the parameters can be varied independently from each other in each layer. We start with the LL equation
\begin{equation}
    \frac{d\vec{m}_{i}}{dt} = -\gamma_i \vec{m}_{i}\times\vec{H}_{\mathrm{eff}_{i}}
    \label{eq:SILLG_layer}
\end{equation}
where $\gamma_i$ is the layer-dependent gyromagnetic ratio. The energy of the system is given by
\begin{equation}
    E = -\vec{H}_{0}\sum_i \vec{m}_i + \sum_{\left<i,j\right>} H_{\mathrm{E}_{ij}} \vec{m}_i\cdot\vec{m}_j - \frac{1}{2}\sum_i H_{\mathrm{a}_i}\left(m_i^{\mathrm{a}}\right)^2 - \frac{1}{2}H_{\mathrm{b}}\sum_i \left(m_i^{\mathrm{b}}\right)^2.
    \label{eq:SIhamiltonian_layerdep}
\end{equation}
We now use explicitly layer-dependent parameters, in particular the interlayer exchange interaction between layers $i$ and $j$, $H_{\mathrm{E_{i,j}}}$, as well as the intermediate axis anisotropy in individual layers $i$, $H_{\mathrm{a}_i}$. The external field and easy-axis anisotropy remain layer-independent. The effective field in layer $i$ then results to 
\begin{equation}
    \vec{H}_{\mathrm{eff}_{i}} = -\nabla_{\vec{m}_{i}} E = \vec{H}_{0} - \sum_{<i,j>} H_{E_{i,j}} \vec{m}_j + H_{\mathrm{a}_i}m_i^{\mathrm{a}}\hat{a} + H_{\mathrm{b}} m_i^{\mathrm{b}}\hat{b}.
    \label{eq:SIefffield_layerres}
\end{equation} 
We can still describe the equilibrium macrospins $\vec{m}_{i,0}$ by their tilt angle $\theta_i$ away from the easy $\hat{b}$-axis: $\vec{m}_{i,0} = \left( 0, \sin \theta_i, \pm \cos \theta_i \right)$. However, the $\theta_i$ are now layer-dependent. By minimizing the energy in \cref{eq:SIhamiltonian_layerdep} with respect to the $\theta_i$, we can find their equilibrium position, analogous to the bulk model in \cref{eq:SIangle_eq}. For three layers, the explicit system of equations is
\small
\begin{align}
    \frac{\mathrm{d}E}{\mathrm{d}\theta_1} &= -H_0 \cos\theta_1 + H_{\mathrm{E}_{1,2}}(\cos\theta_1\sin\theta_2+\sin\theta_1\cos\theta_2)+H_{\mathrm{b}}\cos\theta_1\sin\theta_1 = 0 \\
    \frac{\mathrm{d}E}{\mathrm{d}\theta_2} &= -H_0 \cos\theta_2 + H_{\mathrm{E}_{1,2}}(\cos\theta_2\sin\theta_1+\sin\theta_2\cos\theta_1)+H_{\mathrm{E}_{2,3}}(\cos\theta_2\sin\theta_3+\sin\theta_2\cos\theta_3)+H_{\mathrm{b}}\cos\theta_2\sin\theta_2 = 0 \\
    \frac{\mathrm{d}E}{\mathrm{d}\theta_3} &= -H_0 \cos\theta_3 + H_{\mathrm{E}_{2,3}}(\cos\theta_3\sin\theta_2+\sin\theta_3\cos\theta_2)+H_{\mathrm{b}}\cos\theta_3\sin\theta_3 = 0,
    \label{eq:SItheta_numerical}
\end{align}
\normalsize
which we solve numerically. \\
Next, we again linearize the LL equation in \cref{eq:SILLG_layer} and write $\delta\vec{m}_{i} = \left( \delta m_i^{\mathrm{a}}, \mp \delta m_i^{||} \cos \theta_i, \delta m_i^{||} \sin \theta_i\right)$. Inserting this into \cref{eq:SIlinearizedLLG} and using the ansatz $\delta \vec{m}_i (t) = \delta \vec{m}_i (t=0) e^{-i\omega t}$ results in the following matrix equation for the case of three layers:
\tiny
\begin{equation}
    \setlength{\arraycolsep}{1pt} 
    \medmuskip = 1mu 
    \frac{i\omega}{\gamma} \begin{pmatrix}
        \delta m_1^a \\ \delta m_1^{||} \\ \delta m_2^a \\ \delta m_2^{||} \\ \delta m_3^a \\ \delta m_3^{||}
    \end{pmatrix} = \begin{pmatrix}
        0 & -H_{E_{1,2}} \frac{\cos \theta_2}{\cos\theta_1} -H_{\mathrm{b}}\cos^2 \theta_1 & 0 & H_{E_{1,2}}\cos (\theta_1+\theta_2) & 0 & 0 \\
        H_{E_{1,2}}\frac{\cos \theta_2}{\cos\theta_1} + H_{\mathrm{b}} - H_{a_1} & 0 & H_{E_{1,2}} & 0 & 0 & 0\\
        0 & H_{E_{1,2}} \cos (\theta_1+\theta_2) & 0 & -H_{E_{1,2}} \frac{\cos \theta_1}{\cos\theta_2} -H_{E_{2,3}} \frac{\cos \theta_3}{\cos\theta_2} -H_{\mathrm{b}}\cos^2 \theta_2 & 0 & H_{E_{2,3}}\cos (\theta_2+\theta_3) \\\\
        H_{E_{1,2}} & 0 & H_{E_{1,2}}\frac{\cos \theta_1}{\cos\theta_2} + H_{E_{2,3}}\frac{\cos \theta_3}{\cos\theta_2}  + H_{\mathrm{b}} - H_{a_2} & 0 & H_{E_{2,3}} & 0 \\
        0 & 0 & 0 & H_{E_{2,3}}\cos (\theta_2+\theta_3) & 0 &  -H_{E_{2,3}} \frac{\cos \theta_2}{\cos\theta_3} -H_{\mathrm{b}}\cos^2 \theta_3\\
        0 & 0 & H_{E_{2,3}} & H_{E_{2,3}}\frac{\cos \theta_2}{\cos\theta_3} + H_{\mathrm{b}} - H_{a_3} & 0 
    \end{pmatrix}\begin{pmatrix}
        \delta m_1^a \\ \delta m_1^{||} \\ \delta m_2^a \\ \delta m_2^{||} \\ \delta m_3^a \\ \delta m_3^{||}
    \end{pmatrix}
    \label{eq:SImatrix_numericalmodel}
\end{equation}
\normalsize
which can be solved numerically. \cref{eq:SImatrix_numericalmodel} can be expanded for an arbitrary amount of layers $N$. For large $N$, we extract the $k_{\mathrm{z}}$ value of each eigenmode via Fast Fourier transform and arrive at the dispersion relation for thick samples, as shown in the main text Fig. 3a for $N = 100$.
\section*{Excitation of coherent magnon modes by laser pulses}
\begin{figure*}[h]
\renewcommand\figurename{SI Figure}
    \centering
    \includegraphics[width=\linewidth, trim = 0mm 50mm 0mm 40mm,clip=true]{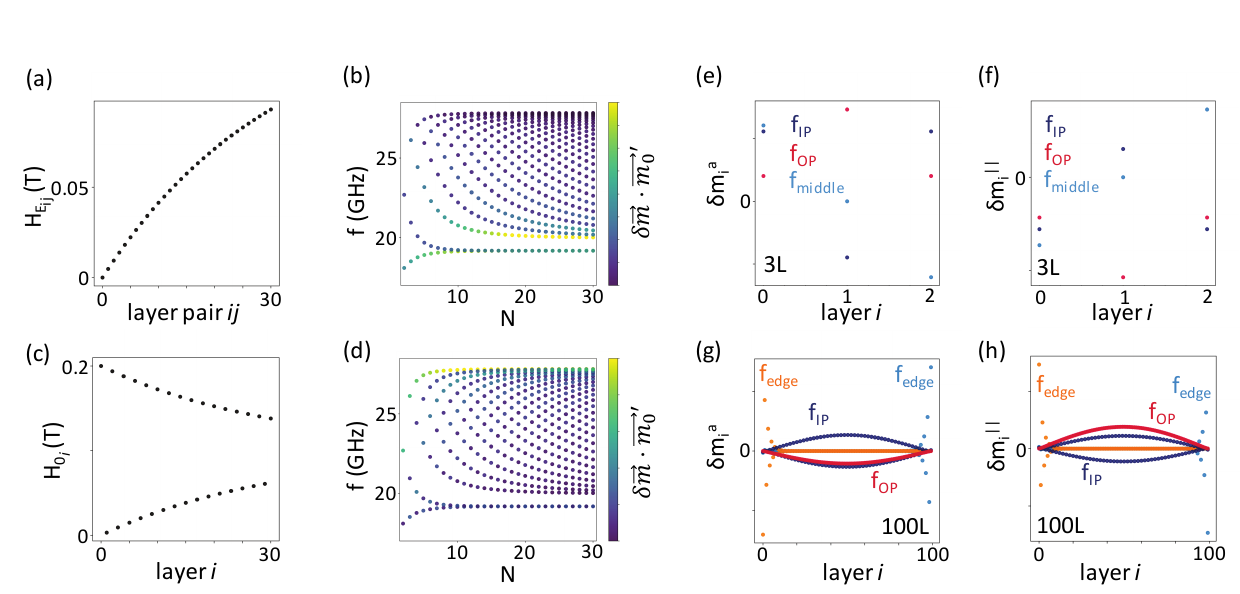}
    \caption{\textbf{Numerical macrospin model} (a) Homogeneous change in $H_{\mathrm{E}}$ across layers (e.g. due to heating) leads to (b) a strong coupling with the acoustic mode for bulk samples or the "surface" mode for thin samples. 
    (c) An alternating change in $H_0$ (which is expected to arise from, e.g., magnetoelastic effects) strongly couples to excitation of the optical mode (d).
    (e-f) Mode shapes of the trilayer.
    (g-h) Surface modes and example of a bulk mode for a 100-layer device.
     }
    \label{fig:SI6}
\end{figure*}

In thick samples, previous studies have established that light pulses excite the optical and acoustic mode around $k_{z} \approx 0$ \cite{Bae2022,Sun2024}. In thin samples, however, the spacing along the z-direction becomes discrete, so that the notion of a wavevector $k_{\mathrm{z}}$ loses meaning. Therefore, we need to establish which modes in the layer-resolved macrospin model correspond to the modes we excite and probe in the experiment in few-layer samples (and to what modes in bulk samples they correspond). We use the following set of assumptions to estimate the coupling of the eigenmodes to excitation by light:
\begin{itemize}
    \item The ultrafast pump pulse leads to a step-like change in the layer-resolved effective fields: 
    \begin{equation}
        \vec{H}_{\mathrm{eff}_j} (t>0) = \vec{H}_{\mathrm{eff}_j}(t<0) + \vec{\Delta H}_{\mathrm{eff}_j}
    \end{equation}
    \item This step-like change can be in the same direction across layers -- e.g. due to heating decreasing the exchange interaction -- or alternate in direction in every second layer -- e.g. due to magnetoelastic coupling \cite{Bae2024}:
    \begin{equation}
        \vec{\Delta H}_{\mathrm{eff}_j} = \vec{\Delta H}_{\mathrm{eff}_{\mathrm{even}}}+ (-1)^j \vec{\Delta H}_{\mathrm{eff}_{\mathrm{odd}}}
    \end{equation}
    \item The change in effective field leads to a new equilibrium macrospin position:
    \begin{equation}
        \vec{m}_{0,j} (t>0) = \vec{m}_{0,j} (t<0) + \Delta \vec{m}_{j}
    \end{equation}
    \item After the excitation, the macrospins want to align along their new equilibrium positions, leading to magnon oscillations. We use the projection of the eigenmodes before the laser pulse $\vec{\delta m}_{j} (t<0)$ onto the $\Delta \vec{m}_{j}$ resulting from the aforementioned excitation as an estimate of how efficiently each eigenmode $\vec{\delta m}_{j} (t<0)$ is excited by light:
    \begin{equation}
        \vec{\delta m}_{j} (t<0) \cdot \Delta \vec{m}_{j} = \vec{\delta m}_{j} (t<0) \cdot (\vec{m}_{0,j}(t>0) - \vec{m}_{0,j}(t<0))
    \end{equation}
    \item We assume that $|m_{j,0}| = const.$, and therefore $\vec{m}_{j,0}(t<0)\cdot\delta\vec{m}_{j}(t<0) = 0$, so that the projection is simply given by 
    \begin{equation}
        \vec{\delta m}_{j}(t<0)\cdot\vec{m}_{0,j}(t>0)
    \end{equation} 
\end{itemize} 
First, we model an excitation which is homogeneous across layers, changing the effective field by 
\begin{equation}
   \vec{\Delta H}_{\mathrm{eff}_{\mathrm{even,i}}} = -\Delta H_{\mathrm{E}_{ij}}\frac{\vec{H}_{\mathrm{E}_{ij}}}{|\vec{H}_{\mathrm{E}_{ij}}|}, 
\end{equation}
i.e. reducing the interlayer exchange in every layer. To account for heating effects, we include a gradient in $\Delta H_{\mathrm{E}_{ij}}$ (resulting layer-dependent exchange interaction after laser excitation in \cref{fig:SI6}a). 
As a result of the decreased exchange interaction, the equilibrium macrospin position after excitation points more along the $\hat{c}$-direction in all layers: 
\begin{equation}
    \vec{m}_{i,0} (t>0) = \vec{m}_{i,0} (t<0) + \Delta m_{i} \hat{c}.
\end{equation} 
Now, we calculate the projection $\vec{\delta m}_{i}(t<0)\cdot\vec{m}_{i,0}(t>0)$ for this case, shown colorcoded in the layer-dependent eigenvalues in \cref{fig:SI6}b. For large $N$, the projection is largest for the lowest frequency bulk modes (yellow dots at $\sim \SI{20}{\giga\hertz}$), which is consistent with the $f_{\mathrm{IP}}$ of bulk measurements \cite{Bae2022}. For small $N$, one of the edge modes starts to couple strongly to light as it spreads across the sample (light green dots with lowest $f$ for $N < 10$), while the coupling of the former $f_{\mathrm{IP}}$ becomes weaker. Therefore we assign the corresponding eigenvalue of this edge mode to the experimental $f_{\mathrm{IP}}$ in few-layer devices.\\
Next, we explore the effect of an alternating laser-pulse-induced change across layers, e.g. an additional effective field term along the c-direction due to magnetoelastic coupling \cite{Bae2024}:
\begin{equation}
    \vec{\Delta H}_{\mathrm{eff}_{\mathrm{odd,j}}} = (-1)^j\Delta H_{\mathrm{T}_j}\hat{c}.
 \end{equation}
As $\vec{H}_0 = H_0\hat{c}$ in our model, we introduce this additional effective field as a change in $H_0$, including a gradient due to heating effects (see \cref{fig:SI6}c).
The new equilibrium macrospin positions after excitation now alternate between layers: 
\begin{equation}
    \vec{m}_{0,j} (t>0) = \vec{m}_{0,j} (t<0) + (-1)^j \Delta m_{j} \hat{c}.
\end{equation}
The resulting projections are shown colorcoded in \cref{fig:SI6}d. The largest projection is seen for the highest frequency mode -- corresponding to the $f_{\mathrm{OP}}$ ($k_{\mathrm{z}} \approx 0$) mode in bulk experiments. For small $N$, the largest projection also belongs to the mode with the largest eigenvalue, which we therefore assign to our experimental $f_{\mathrm{OP}}$ (at $N=3$ the bulk $f_{\mathrm{IP}}$ and $f_{\mathrm{OP}}$ modes converge). In Fig. 3b of the main text, the modes are colored according to the coupling to a homogeneous (blue) vs. alternating (red) excitation. \\
In \cref{fig:SI6}e, f we show the components $m_a$ and $m_{||}$ for the trilayer modes. The mode $f_{\mathrm{middle}}$ (bright blue) is asymmetric across the device, while $f_{\mathrm{IP}}$ (dark blue) and $f_{\mathrm{OP}}$ (red) are symmetric. An excitation by light should have a similar effect in every second layer (due to the long wavelength of light compared to the sample thickness), and therefore cannot couple to the $f_{\mathrm{middle}}$ mode.
In \cref{fig:SI6}g, h we show the $f_{\mathrm{IP}}$ (dark blue) and $f_{\mathrm{OP}}$ (red) and the two edge modes (bright blue, green) in a 100-layer device. The edges modes are confined to the first $\sim 8$ layers. The $f_{\mathrm{IP}}$ and $f_{\mathrm{OP}}$ modes spread across the complete bulk and have the maximum possible real space wavelength, i.e. minimal $k_{\mathrm{z}}$, which is consistent with the argumentation in previous studies that light excitation couples to these two modes.
\newpage
\section*{Macrospin fit results}
\begin{figure*}[h]
\renewcommand\figurename{SI Figure}
    \centering
    \includegraphics[width=\linewidth, trim = 0mm 50mm 0mm 50mm,clip=true]{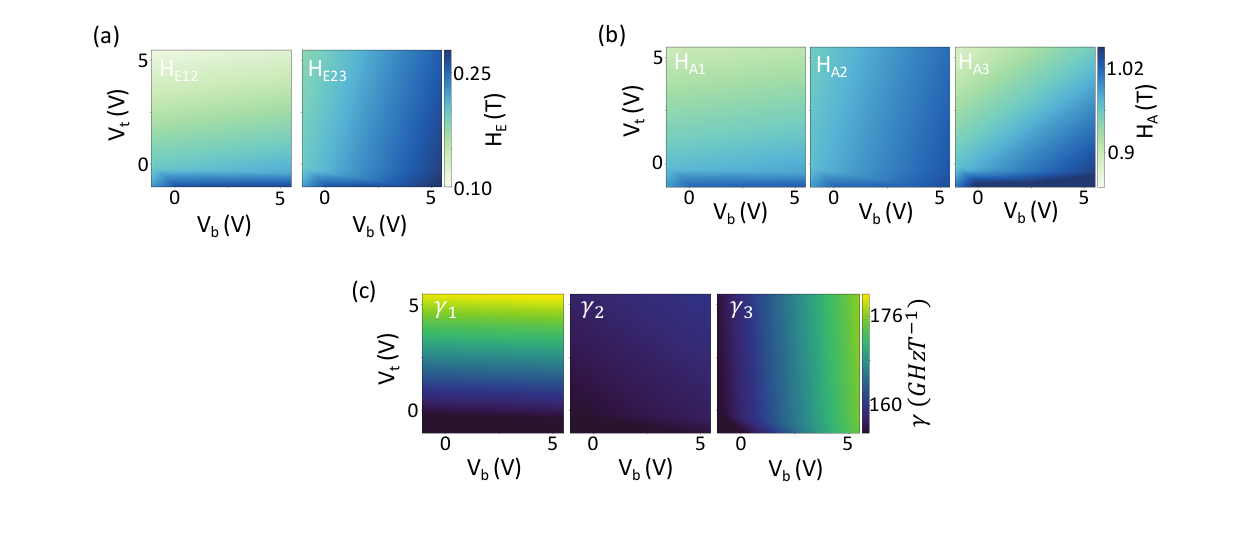}
    \caption{\textbf{Macrospin fitting results trilayer} The results of the macrospin model fitted to the capacitor model for the (a) exchange fields, (b) anisotropy fields and (c) gyromagnetic ratios.
     }
    \label{fig:SI7}
\end{figure*}
\begin{figure*}[h!]
\renewcommand\figurename{SI Figure}
    \centering
    \includegraphics[width=\linewidth, trim = 0mm 35mm 0mm 40mm,clip=true]{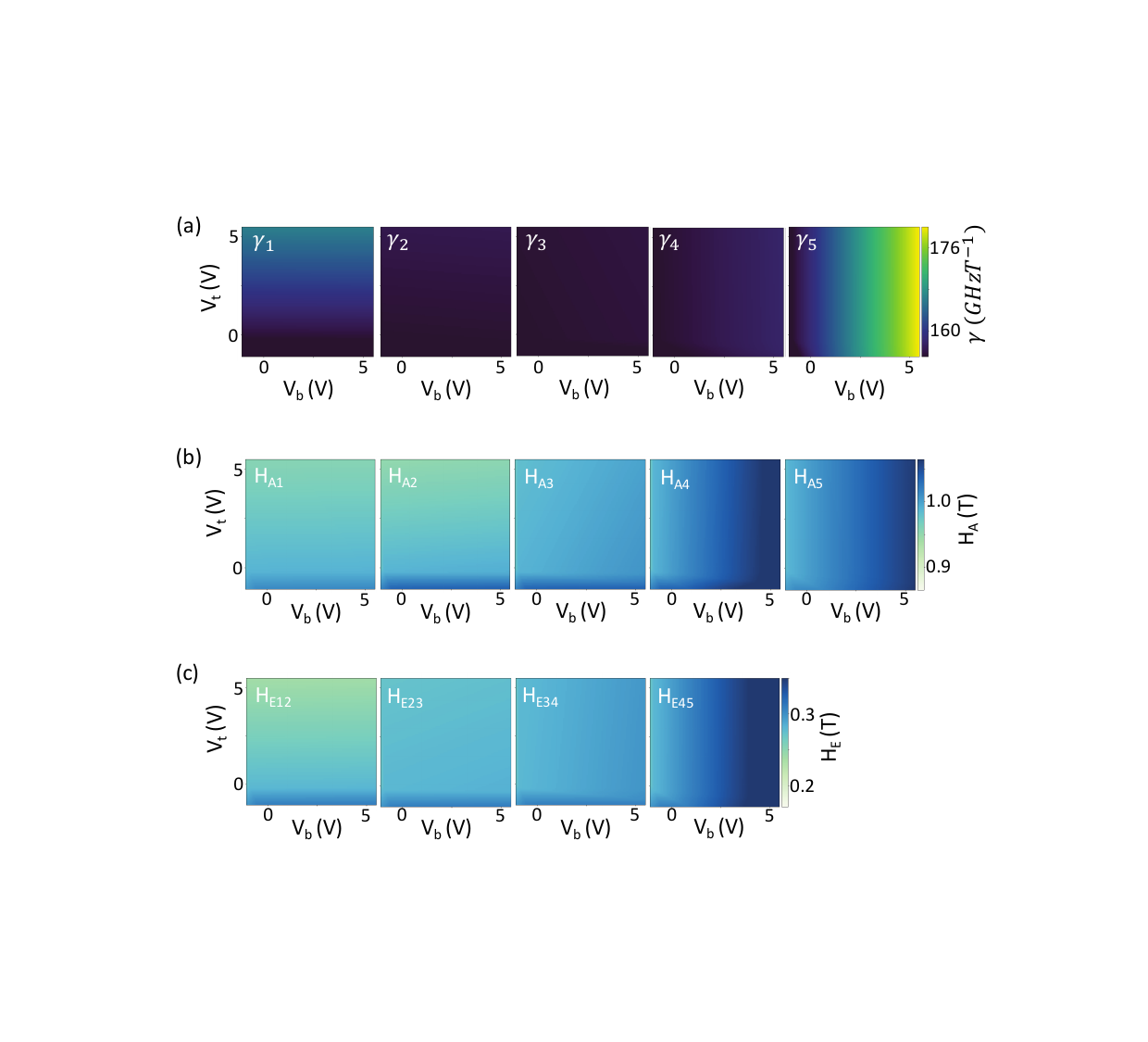}
    \caption{\textbf{Macrospin fitting results 5-layer} The results of the macrospin model fitted to the capacitor model for the (a) gyromagnetic ratios, (b) anisotropy fields and (c) interlayer exchange fields.
     }
    \label{fig:SI71}
\end{figure*}

\begin{figure*}[h!]
\renewcommand\figurename{SI Figure}
    \centering
    \includegraphics[width=\linewidth, trim = 0mm 30mm 0mm 30mm,clip=true]{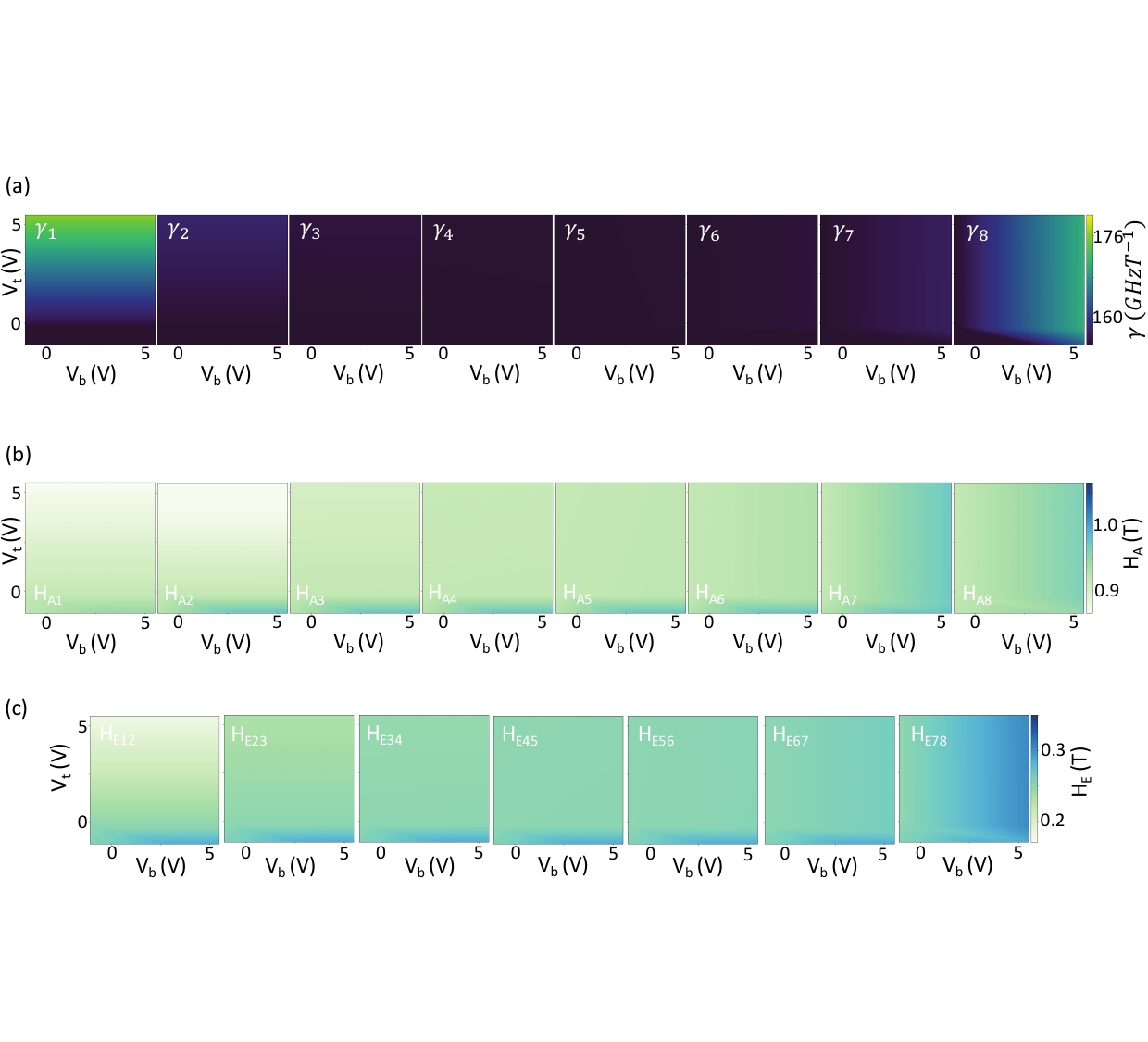}
    \caption{\textbf{Macrospin fitting results 8-layer} The results of the macrospin model fitted to the capacitor model for the (a) gyromagnetic ratios, (b) anisotropy fields and (c) interlayer exchange fields.
     }
    \label{fig:SI72}
\end{figure*}

 
Using the dependencies from the main text and the results from \cref{fig:SI3}, we fit the lowest and highest eigenvalue of \cref{eq:SImatrix_numericalmodel} to the observed gate-dependent $f_{\mathrm{IP}}$ and $f_{\mathrm{OP}}$ of the trilayer. We use the resulting fitting parameters from Table 1 of the main text to calculate the interlayer exchange and intermediate axis anisotropy fields and gyromagnetic ratios for each layer in \cref{fig:SI7}. \\
The maximum changes in gyromagnetic ratio we predict in the top layer roughly correspond to a change of 2.75 to 3.2 $\mu_B$ per Cr atom, which is an order of magnitude more than predicted for CrSeBr by doping \cite{Han2024}.
Our dependence of anisotropy on electric field of $-\SI{0.8}{\tesla\nano\metre\per\volt}$ corresponds to $\sim -\SI{600}{\femto\joule\per\volt\per\metre}$ of changes in anisotropy constant per unit surface per unit electric field. For comparison, usual voltage controlled magnetic anisotropy (VCMA) coefficients describing changes of interfacial perpendicular magnetic aniostropy at ferromagnet-oxide interfaces reach around $\sim -\SI{100}{\femto\joule\per\volt\per\metre}$ \cite{Rana2019,Li2017,Nozaki2013}. The tunability we reach in the interlayer exchange field is comparable to that in bulk CrSBr when applying in-plane strain \cite{Diederich2022}, however, the exchange interaction becomes stronger rather than weaker.\\
We do the same calculations for the 5- and 8-layer devices using the results of the electrostatic modelin in \cref{fig:SI4,fig:SI5} and the fitting parameters from Table 1 of the main text. The resulting internal fields and gyromagnetic ratios are shown in \cref{fig:SI71,fig:SI72}.
\newpage
\section*{Gate dependence of magnons in 8-layer device}
\begin{figure*}[h!]
\renewcommand\figurename{SI Figure}
    \centering
    \includegraphics[width=\linewidth, trim = 0mm 60mm 0mm 50mm,clip=true]{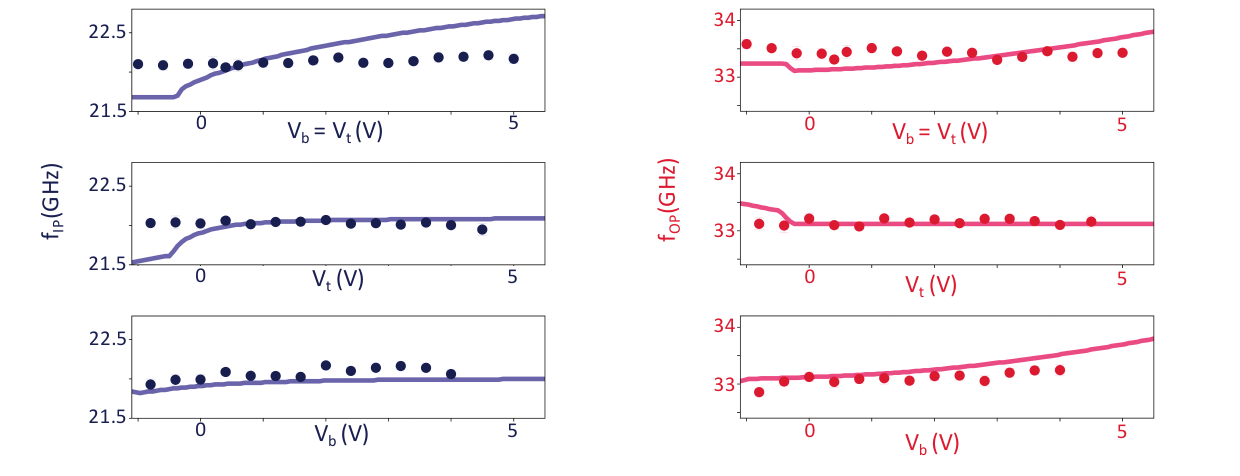}
    \caption{\textbf{Gate-dependent magnon frequencies of 8-layer device} Complete gate-dependence of (a) $f_{\mathrm{IP}}$ and (b) $f_{\mathrm{OP}}$ in the 8-layer device (dots: experimental data, solid lines: macrospin model).
     }
    \label{fig:SI8}
\end{figure*}

The compelete gate-dependence of the magnons in 8 layers is shown in \cref{fig:SI8} together with the macrospin modeling. Overall, the model predicts the experimentally observed frequecies. There are discrepancies mainly in the doping dependence of $f_{\mathrm{IP}}$ (top left panel). These could be due to some simplifications we made for fitting: For example, we fixed $H_b = \SI{1.3}{\tesla}$ in all devices. However, $H_b$ could vary slightly between edgexfoliated flakes -- this can lead to differences in $\nu_{\mathrm{a}}$ across samples. This could result in over- or underestimating changes in $f_{\mathrm{IP}}$ for certain field- and doping conditions. This is reflected in the larger errors for the 8-layer fitting results in Table 1 of the main text.
\section*{Gate-dependent magnetic phenomena}
\begin{figure*}[h!]
\renewcommand\figurename{SI Figure}
    \centering
    \includegraphics[width=0.4\linewidth, trim = 60mm 95mm 105mm 40mm,clip=true]{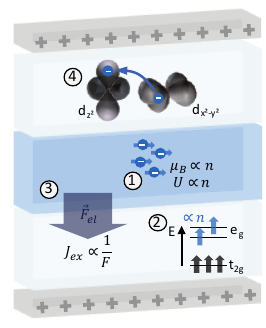}
    \caption{\textbf{Possible doping- and field related effects on magnetic parameters} Sketch of possible gating effects on gyromagnetic ratio, anisotropy and exchange fields.
     }
    \label{fig:SI13}
\end{figure*}

As mentioned in the main text, multiple gating effects on magnetic properties have been suggested as illustrated in \cref{fig:SI13}. (1) By increasing the
electron doping n, the magnitude of the gyromagnetic ratio $\gamma$ of the macrospins can increase due to an increase in carriers with magnetic moment \cite{Han2024}. The Coulomb repulsion parameter U can increase with $n$, possibly influencing the exchange constants according to the Kugel-Khomskii model \cite{Farsane2023,Kugel1973,Kugel1982}. (2) When doping with electrons, $e_g$ levels will be populated in addition to $t_{2g}$ ones \cite{Kugel1973,Kugel1982,Mazurenko2006}. This can change the orbitals involved in the hopping between sites, again affecting the (anti-)ferromagnetic exchange coupling parameters. Additionally, the different
orbital shapes can influence the anisotropy of the material \cite{Ziebel2024}. 
(3) A perpendicular electric field between layers can shift the energy positons of bands associated with different orbitals in neighboring layers with respect to each other, again influencing the (anti-)ferromagnetic exchange coupling parameter \cite{Wang2023_2}. (4) Gates can influence orbital filling especially in the outer layers due to the resulting electric field, as electrons want to reside further/closer to the gate, shuffling them into $d_{x^2-y^2}$ or $d_{z^2}$ orbitals, respectively. This is commonly called voltage-induced magnetic anisotropy \cite{Maruyama2009,He2011,Rana2019}.
\newpage
\section*{Thickness measurement of CrSBr flakes}
\begin{figure*}[h!]
\renewcommand\figurename{SI Figure}
    \centering
    \includegraphics[width=\linewidth, trim = 0mm 45mm 0mm 65mm,clip=true]{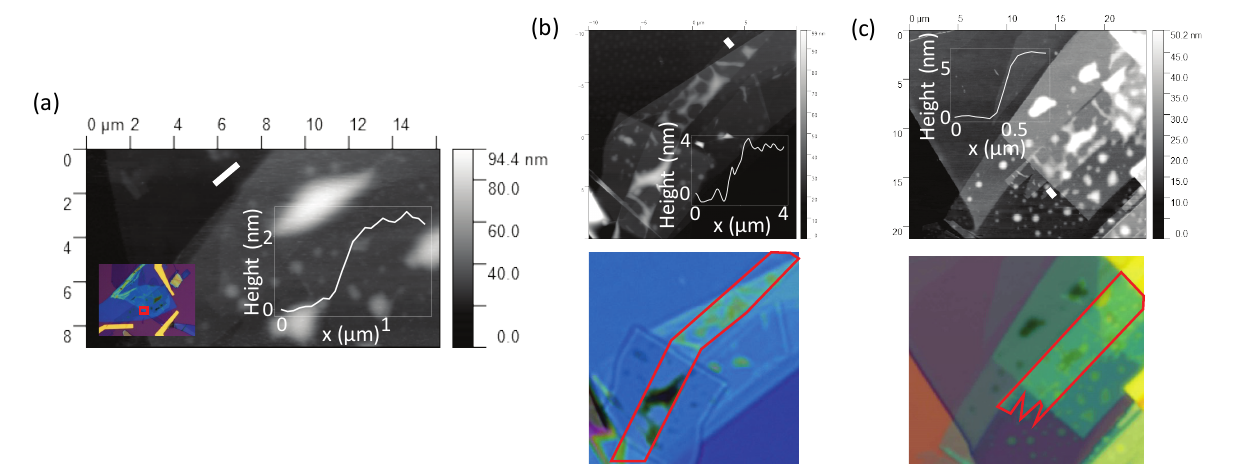}
    \caption{\textbf{AFM images} (a) AFM image of a small portion of the trilayer device (marked in optical image). The height profile corresponds to a linecut of the CrSBr flake along the marker in the image.
    (b) AFM (top) and corresponding optical (bottom, CrSBr flake outlined) image of the 5-layer device before contact patterning. The height profile corresponds to a linecut of the CrSBr flake along the marker in the image.
    (c) AFM (top) and corresponding optical (bottom, CrSBr flake outlined) image of the 8-layer device before contact patterning. The height profile corresponds to a linecut of the CrSBr flake along the marker in the image.
     }
    \label{fig:SI9}
\end{figure*}

To find the layer number of the measured flakes, we performed AFM measurements (\cref{fig:SI9}). We find the following thicknesses -- Sample 1: $\SI{2.4}{\nano\metre}$, Sample 2: $\SI{4}{\nano\metre}$, Sample 1: $\SI{6.4}{\nano\metre}$ -- corresponding to 3, 5 and 8 layers, respectively \cite{Telford2022}.
